\documentclass[pdflatex,sn-mathphys-num,referee]{sn-jnl}%

\usepackage{graphicx}%
\usepackage{multirow}%
\usepackage{amsmath,amssymb,amsfonts}%
\usepackage{amsthm}%
\usepackage{mathrsfs}%
\usepackage[title]{appendix}%
\usepackage{xcolor}%
\usepackage{textcomp}%
\usepackage{manyfoot}%
\usepackage{booktabs}%
\usepackage{algorithm}%
\usepackage{algorithmicx}%
\usepackage{algpseudocode}%
\usepackage{listings}%
\DeclareMathOperator*{\argmax}{arg\,max}

\begin{document}

\title[Article Title]{BEACON - Automated Aberration Correction for Scanning Transmission Electron Microscopy using Bayesian Optimization}

\author[1]{\fnm{Alexander J.} \sur{Pattison}}\email{ajpattison@lbl.gov}
\author[1]{\fnm{Stephanie M.} \sur{Ribet}}\email{sribet@lbl.gov}
\author[2]{\fnm{Marcus M.} \sur{Noack}}\email{mmnoack@lbl.gov}
\author[3,1]{\fnm{Georgios} \sur{Varnavides}}\email{gvarnavides@lbl.gov}
\author[4]{\fnm{Kunwoo} \sur{Park}}\email{pcm403@snu.ac.kr}
\author[5]{\fnm{Earl} \sur{Kirkland}}\email{ejk14@cornell.edu}
\author[4,6,7,8]{\fnm{Jungwon} \sur{Park}}\email{jungwonpark@snu.ac.kr}
\author[9,10]{\fnm{Colin} \sur{Ophus}}\email{cophus@stanford.edu}
\author*[1]{\fnm{Peter} \sur{Ercius}}\email{percius@lbl.gov}

\affil*[1]{\orgdiv{Molecular Foundry}, \orgname{Lawrence Berkeley National Laboratory}, \orgaddress{\street{1 Cyclotron Road}, \city{Berkeley}, \state{California}, \country{USA}, \postcode{94720}}}

\affil[2]{\orgname{Lawrence Berkeley National Laboratory}, \orgaddress{\street{1 Cyclotron Road}, \city{Berkeley}, \state{California}, \country{USA}, \postcode{94720}}}

\affil[3]{\orgdiv{Miller Institute for Basic Research in Science}, \orgname{University of California}, \orgaddress{{\city{Berkeley}, \state{California}, \country{USA}, \postcode{94720}}}}

\affil[4]{\orgdiv{School of Chemical and Biological Engineering, Institute of Chemical Processes}, \orgname{Seoul National University}, \orgaddress{\city{Seoul}, \country{Republic of Korea}, \postcode{08826}}}

\affil[5]{\orgdiv{School of Applied and Engineering Physics}, \orgname{Cornell University}, \orgaddress{\city{Ithaca}, \state{New York}, \country{USA}, \postcode{14853}}}

\affil[6]{\orgdiv{Center for Nanocrystal Research}, 
\orgname{Institute for Basic Science (IBS)}, \orgaddress{\city{Seoul}, \country{Republic of Korea}, \postcode{08826}}}

\affil[7]{\orgdiv{Institute of Engineering Research, College of Engineering}, \orgname{Seoul National University}, \orgaddress{\city{Seoul}, \country{Republic of Korea}, \postcode{08826}}}

\affil[8]{\orgdiv{Advanced Institute of Convergence Technology}, \orgname{Seoul National University}, \orgaddress{\city{Suwon}, \country{Republic of Korea}, \postcode{16229}}}

\affil[9]{\orgdiv{Department of Materials Science and Engineering}, \orgname{Stanford University}, \orgaddress{\city{Stanford}, \state{California}, \country{USA}, \postcode{94305}}}

\affil[10]{\orgdiv{Precourt Institute for Energy}, \orgname{Stanford University}, \orgaddress{\city{Stanford}, \country{USA}, \postcode{94305}}}

\abstract{Aberration correction is an important aspect of modern high-resolution scanning transmission electron microscopy.
Most methods of aligning aberration correctors require specialized sample regions and are unsuitable for fine-tuning aberrations without interrupting on-going experiments. 
Here, we present an automated method of correcting first- and second-order aberrations called BEACON  which uses Bayesian optimization of the normalized image variance to efficiently determine the optimal corrector settings. 
We demonstrate its use on gold nanoparticles and a hafnium dioxide thin film showing its versatility in nano- and atomic-scale experiments. 
BEACON can correct all first- and second-order aberrations simultaneously to achieve an initial alignment and first- and second-order aberrations independently for fine alignment. 
Ptychographic reconstructions are used to demonstrate an improvement in probe shape and a reduction in the target aberration.}


\maketitle

\section{Introduction}\label{sec1}

Aberration correction is an important component of high-resolution scanning transmission electron microscopy (STEM). Modern multipole aberration correctors effectively correct for the main spherical aberration (C\textsubscript{3} in round electron lenses) but add smaller parasitic aberrations that also need to be corrected~\cite{kirkland2018fine,biskupek2012effects,haider2000upper,martis2023imaging,li2020ten}. This is typically accomplished using an automated program based on the analysis of Ronchigrams (for example \cite{Lupini2011ronchi}) or a (CEOS) probe tableau to estimate aberration coefficients from a set of images at different defocii and beam tilts. The program then modifies multipole lenses to compensate for the measured aberrations~\cite{krivanek1999towards,haider1998electron,zemlin1978coma,bertoni2023near}. Recognizable features in Ronchigrams (such as streaks) can also be used to manually correct lower-order aberrations such as 2-fold astigmatism (A\textsubscript{1}) and axial coma (B\textsubscript{2}). Microscope and corrector lens settings drift on the minutes to hours time scales and fine-tuning during an experiment is necessary to achieve optimal performance~\cite{barthel2013optical}. Methods based on the Ronchigram require a flat amorphous sample region (e.g. a thin carbon film), which can be difficult to find on monolithic crystalline films. For the probe tableau method, a special sample is required and is typically only used to tune the corrector at the start of an experiment. A general method for tuning aberrations that can be used on a wide range of samples is therefore desirable. Ideally, such a method would be fully automated and apply the least amount of beam dose possible to minimize time, damage (especially to beam-sensitive samples), and carbon contamination of a region of interest.

Recently, there has been promising progress in the development of new methods for automating aberration correction. DeepFocus uses a convolutional neural network to calculate the defocus and astigmatism in STEM images from two images at different focus values~\cite{schubert2024deepfocus}. A convolutional neural network was used to predict beam emittance from simulated Ronchigrams and iteratively compensate for first- and second-order aberrations~\cite{ma2023physics}. Another convolutional neural network was also used to directly assess aberrations from a single Ronchigram and iteratively correct them~\cite{bertoni2023near}. In Ref.~\citenum{kirkland2018fine}, Kirkland demonstrated via simulation that the normalized variance ($\sigma^2/\mu^2$) of a high-angle annular dark field (HAADF-) STEM image increases when the magnitude of first, second and third-order aberrations decreases. Ishikawa et al. experimentally demonstrated the same relationship using image standard deviation ($\sigma$) to correct first- and second-order aberrations~\cite{ishikawa2021automated}. Together, these works demonstrate that adjusting aberrations to maximize $\sigma$ or $\sigma^2/\mu^2$ can effectively correct aberrations without directly measuring them. Kirkland used a simplex method to iteratively determine the set of aberration coefficients that maximizes $\sigma/\mu^2$ using STEM image simulations~\cite{kirkland2018fine}, while Ishikawa et al. developed a system that maximized $\sigma$ for each aberration coefficient sequentially~\cite{ishikawa2021automated}. A similar concept is utilized by companies such as Thermo Fischer in their OptiSTEM+ software and has been shown to improve microscope alignment for atomic resolution imaging.

In this paper, we present BEACON --- Bayesian-Enhanced Aberration Correction and Optimization Network --- which uses Bayesian optimization to autonomously correct first- and second-order aberrations, both individually and collectively.
Bayesian optimization can efficiently estimate unknown functions with a reduced set of measurements, making it useful in situations where each measurement is expensive in terms of time, cost and/or, in this case, beam dose.
It is used for autonomous experiments in synchrotrons~\cite{noack2021gaussian}, fMRI studies~\cite{lorenz2017neuroadaptive} and scanning-probe microscopy~\cite{thomas2022autonomous}. 
We employ the metric of normalized variance (NV), proposed by Kirkland for HAADF-STEM and others for SEM, as a metric for image quality~\cite{erasmus1982automatic,santos1997evaluation, rudnaya2010evaluating, dembele2016combining}. We demonstrate the use of BEACON at atomic- and nanoscale resolution using a polycrystalline hafnium dioxide (HfO\textsubscript{2}) thin film and a sample composed of 5 nm gold nanoparticles (Au NP) on a carbon support to show the method's viability in different experimental scenarios. We demonstrate that BEACON efficiently compensates for aberrations, leading to higher quality and higher resolution images. Further, we show through ptychographic reconstruction of pre- and post-corrected data that BEACON indeed reduces the target aberrations. The simplicity and robustness of BEACON makes it useful for fine-tuning aberration corrector alignment in low- and high-resolution experiments without human intervention.

\section{Results}\label{sec2}

\subsection{Verifying image NV relationship}\label{subsecNV}

\begin{figure}
\centering
\includegraphics [width=\textwidth] {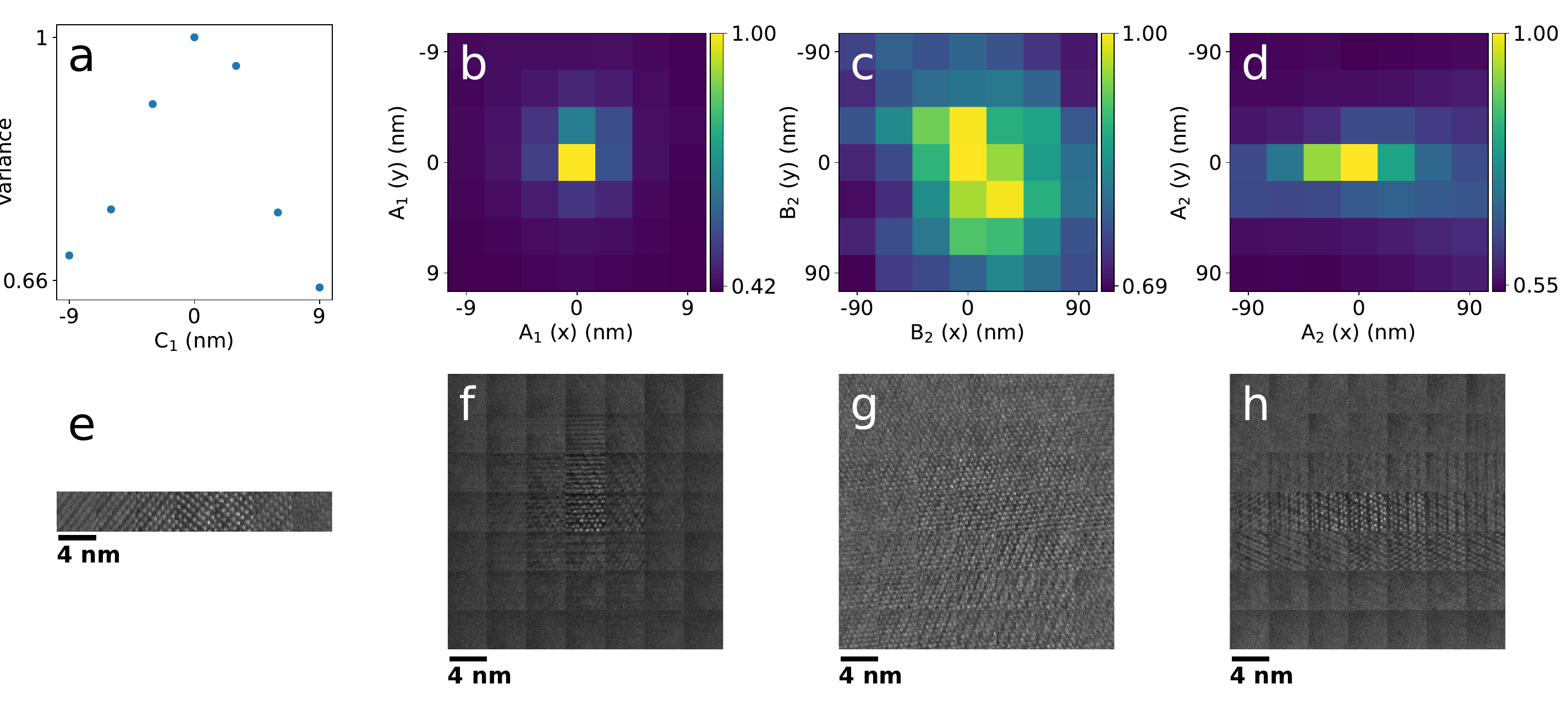}
\caption{Grid searches of C\textsubscript{1} (a, e), A\textsubscript{1} (b, f), B\textsubscript{2} (c, g) and A\textsubscript{2} (d, h) for the HfO\textsubscript{2} thin film. a-d) Image NV normalized to the maximum value. e-h) Tiled cutouts (4.3 nm field of view) of the corresponding full real space images. The zero aberration value was the initial well-corrected state as determined by the CEOS alignment software.}
\label{fig:HfO2 Grid Search}
\end{figure}

\begin{figure}
\centering
\includegraphics [width=\textwidth] {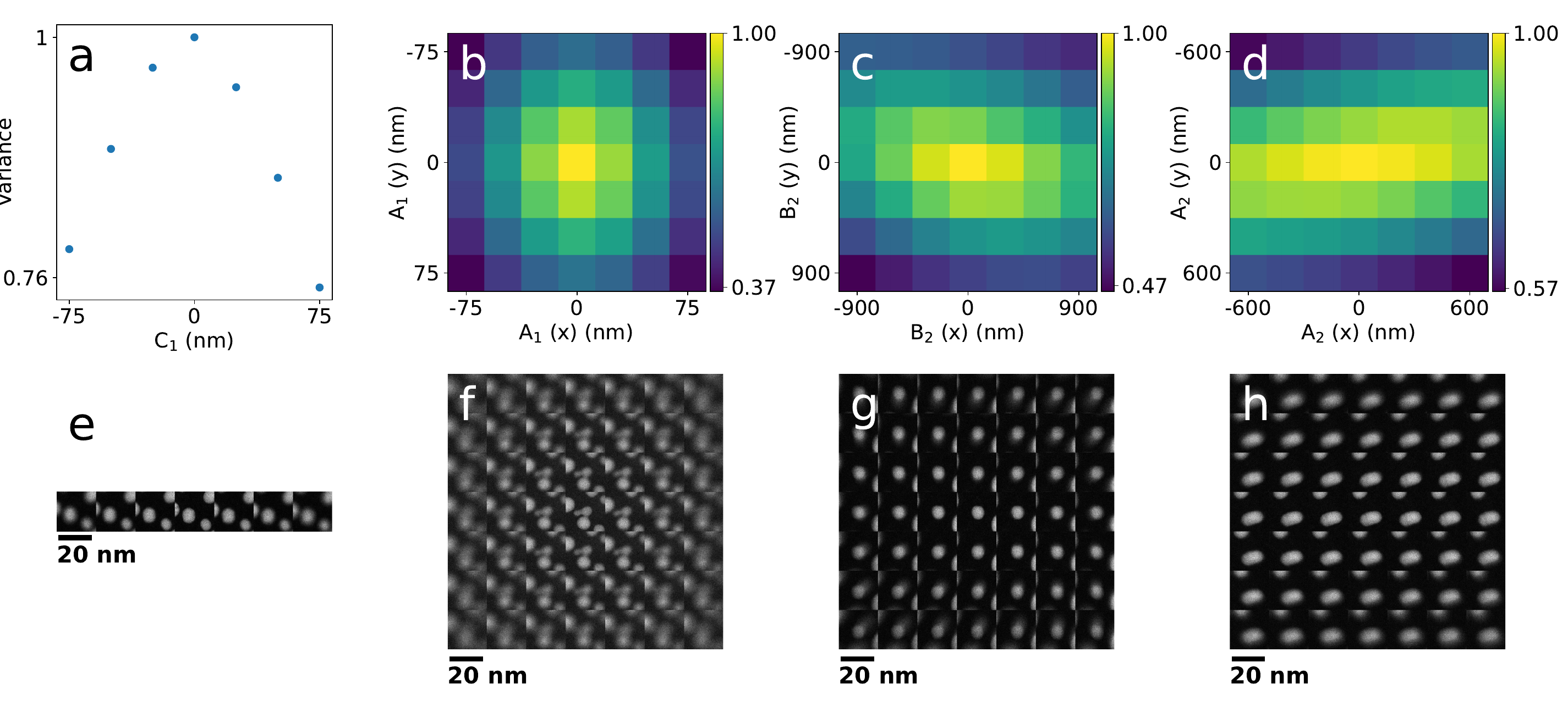}
\caption{Grid searches of C\textsubscript{1} (a, e), A\textsubscript{1} (b, f), B\textsubscript{2} (c, g) and A\textsubscript{2} (d, h) for the Au NP sample. a-d) Image NV normalized to the maximum value. e-h) Tiled cutouts (24.4 nm field of view) of the corresponding full real space images. The zero aberration value was the initial well-corrected state as determined by the CEOS alignment software.}
\label{fig:AuNP Grid Search}
\end{figure}

To verify the correlation between the image NV and aberration coefficient magnitude for both samples, we systematically varied corrector aberration values in small steps (called a ``grid search'') and acquired HAADF-STEM images for each of the first- and second-order aberrations: defocus (C\textsubscript{1}), two-fold astigmatism (A\textsubscript{1}), axial coma (B\textsubscript{2}), and three-fold astigmatism (A\textsubscript{2}). Figure \ref{fig:HfO2 Grid Search} (HfO\textsubscript{2} thin film) and Figure \ref{fig:AuNP Grid Search} (Au NPs) demonstrate that the image NV (top row) and visual image quality (bottom row) both vary inversely with the aberration coefficient magnitude~\cite{kirkland1990image, ishikawa2021automated}.

\subsection{Individual Aberrations}

\begin{figure}
\centering
\includegraphics [width=\textwidth] {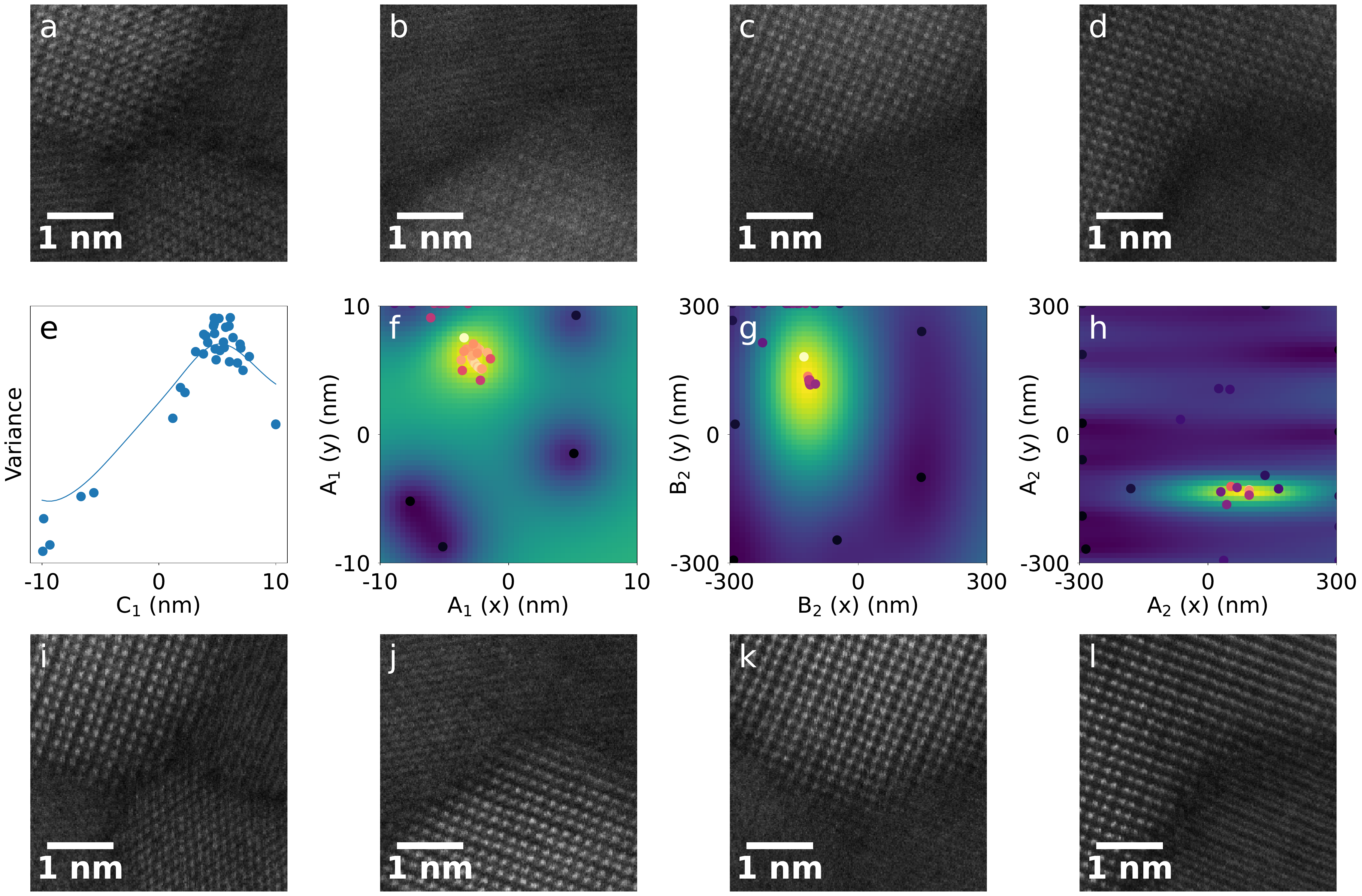}
\caption{BEACON corrections of C\textsubscript{1} (offset = 5 nm) (a, e, i), A\textsubscript{1} (x offset = -5 nm, y offset = 5 nm) (b, f, j), B\textsubscript{2} (x offset = -150 nm, y offset = 150 nm) (c, g, k), and A\textsubscript{2} (x offset = 100 nm, y offset = -150 nm) (d, h, l) aberrations on HfO\textsubscript{2} sample. a-d) Images before correction. e-h) Surrogate models with sampled points marked by dots. i-l) Images after correction.}
\label{fig:HfO2 gpCAM}
\end{figure}

We initially tested BEACON on each first- and second-order aberration individually. 
We started from a corrected state before applying a known aberration value and running BEACON for a total of 35 iterations. 
Unless otherwise specified, we always initialized the model with 5 randomly chosen points followed by a specified number of iterations using the upper confidence bound (UCB) algorithm (see Section \ref{BO}).
Images were taken before and after the runs to verify the results.
The middle row of Figure \ref{fig:HfO2 gpCAM} shows the final model overlaid with each experimental measurement (colored dots). BEACON finds the peak of the surrogate model (background image), while the top and bottom rows show pre- and post-correction HAADF-STEM images of the HfO\textsubscript{2} thin film, respectively, and demonstrate a visual improvement in the image quality. 
Supplementary Figure \ref{fig:AuNP gpCAM} shows the same results for the Au NP sample.

\begin{figure}
\centering
\includegraphics [width=\textwidth] {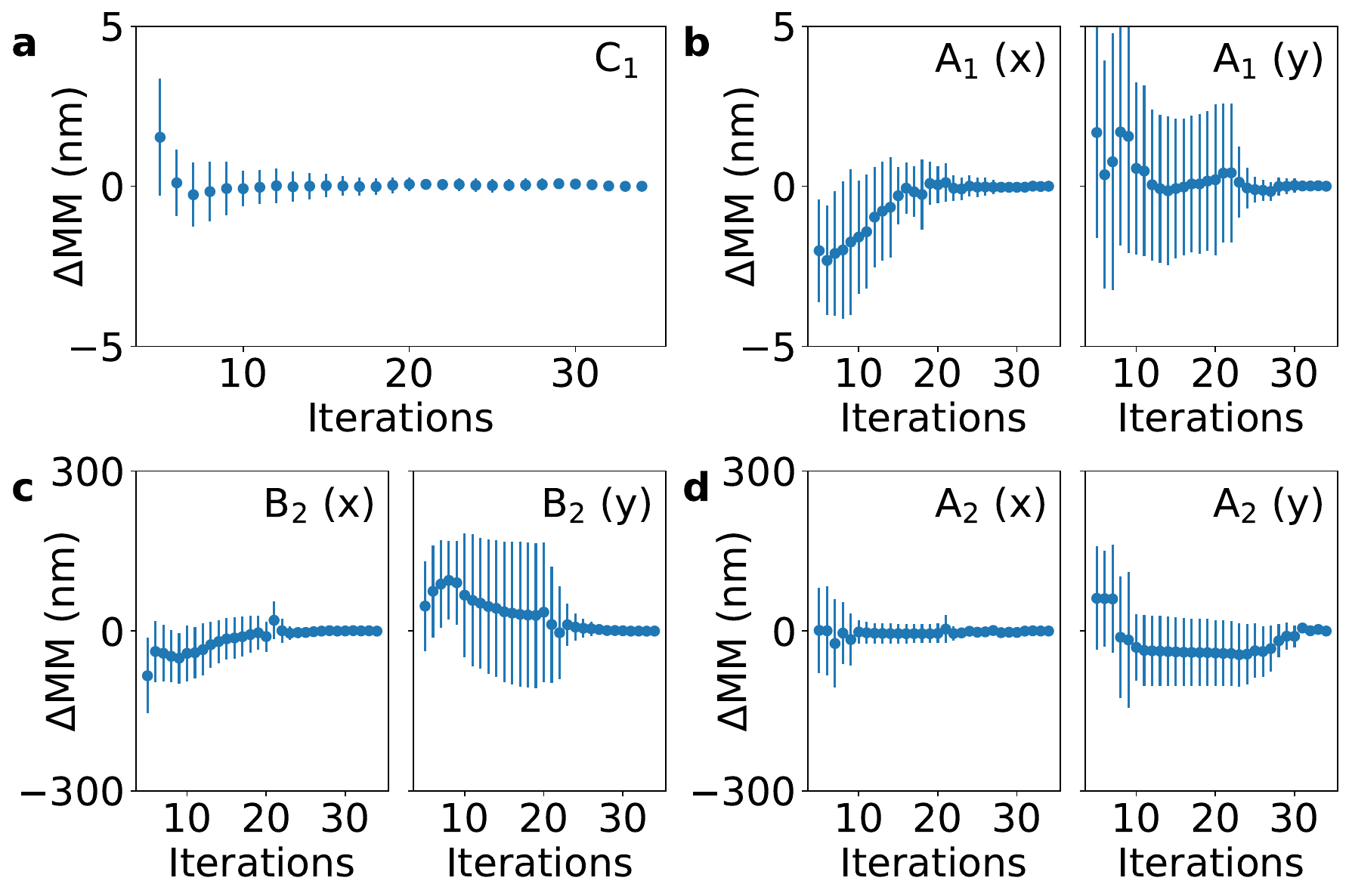}
\caption{Plots of the average difference between the model maximum and the final model maximum ($\Delta$MM) during BEACON runs on the HfO\textsubscript{2} thin film for a) C\textsubscript{1}, b) A\textsubscript{1}, c) B\textsubscript{2} and d) A\textsubscript{2}.}
\label{fig:individual convergences HfO2}
\end{figure}

\begin{figure}
\centering
\includegraphics [width=\textwidth] {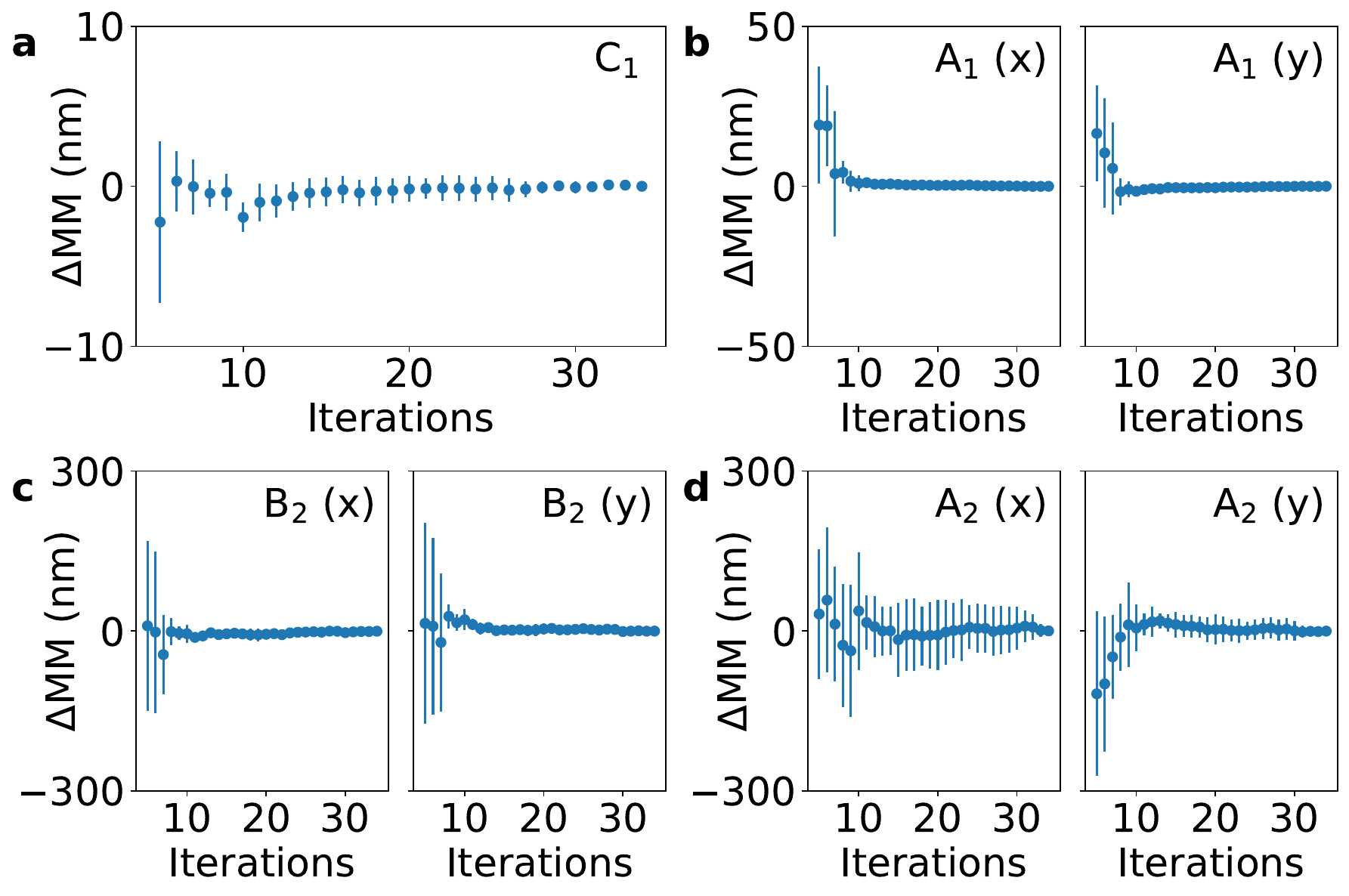}
\caption{Plots of the average difference between model maximum and final model maximum ($\Delta$MM) during BEACON runs on the Au NP sample for a) C\textsubscript{1}, b) A\textsubscript{1}, c) B\textsubscript{2} and d) A\textsubscript{2}.}
\label{fig:individual convergences AuNP}
\end{figure}

To quantitatively assess the performance and reliability of BEACON, we did 5 runs on each sample starting from the same initial aberration settings. 
We evaluated the correction precision by calculating the standard error of all final aberration coefficients compared to the mean, as a measure of self-consistency (see Table \ref{tab:individual statistics}). 
We also assessed the convergence of BEACON by measuring the difference between the final maximum of the surrogate model and the maximum of the surrogate model after every iteration, then calculating the standard deviations of these differences over the 5 runs (see Figure \ref{fig:individual convergences HfO2} and Supplementary Figure \ref{fig:individual convergences AuNP}). 
We set benchmarks for a good corrector alignment as $\pm 1$ nm for first-order corrections (C\textsubscript{1}, A\textsubscript{1}) and $\pm 20$ nm for second-order corrections (B\textsubscript{2}, A\textsubscript{2}). 
We defined the number of iterations needed for convergence as the iteration at which the standard deviation of the differences fell below these benchmarks (see Table \ref{tab:individual statistics}).

\begin{table}
\centering
\begin{tabular}{|c|c|c|c|c|}
\hline
\begin{tabular}[c]{@{}c@{}}Corrected\\ aberration\end{tabular} & \begin{tabular}[c]{@{}c@{}}Precision of HfO$_2$\\ measurement (nm)\end{tabular} & \begin{tabular}[c]{@{}c@{}}Convergence \\ iteration\\ for HfO$_2$\end{tabular} & \begin{tabular}[c]{@{}c@{}}Precision of Au NP\\ measurement (nm)\end{tabular} & \begin{tabular}[c]{@{}c@{}}Convergence \\ iteration\\ for Au NP\end{tabular} \\ \hline
C\textsubscript{1} & 0.29 & 8 & 1.25 & 8 \\ \hline
A\textsubscript{1} & 1.18 & 24 & 0.17 & 13 \\ \hline
B\textsubscript{2} & 16.18 & 25 & 15.04 & 9 \\ \hline
A\textsubscript{2} & 24.27 & 31 & 40.14 & 9 \\ \hline
\end{tabular}
\caption{Correction precision and number of iterations required for convergence for individually corrected aberrations.}
\label{tab:individual statistics}
\end{table}

\subsection{Combined aberrations}
In the previous section, we corrected each aberration separately with a maximum of two dimensions. 
However, Bayesian optimization can handle higher dimensional problems and each image NV provides information about multiple aberrations simultaneously.

\begin{figure}
\centering
\includegraphics [width=\textwidth] {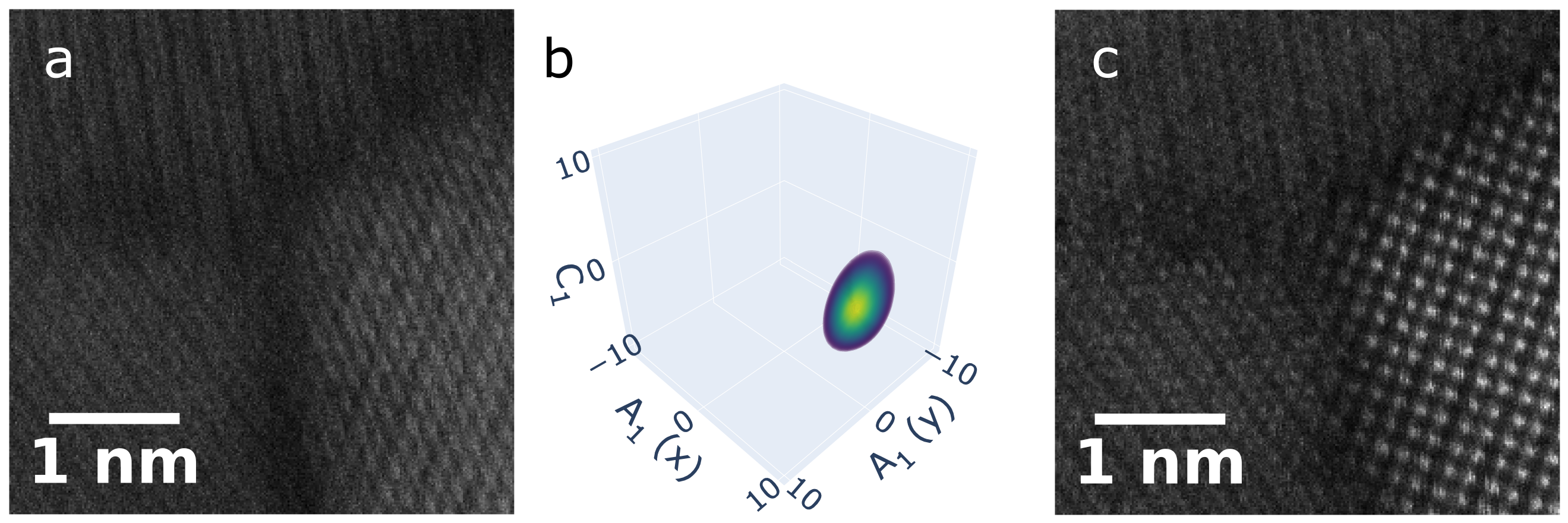}
\caption{BEACON correction of combined C\textsubscript{1} (offset = -5 nm) and A\textsubscript{1} (x offset = 5 nm, y offset = -5 nm) on the HfO\textsubscript{2} thin film. Magnified portion of the full image a) before correction and c) after correction. b) Volumetric plot of the surrogate model.}
\label{fig:C1 A1 gpCAM}
\end{figure}

\begin{figure}
\centering
\includegraphics [width=\textwidth] {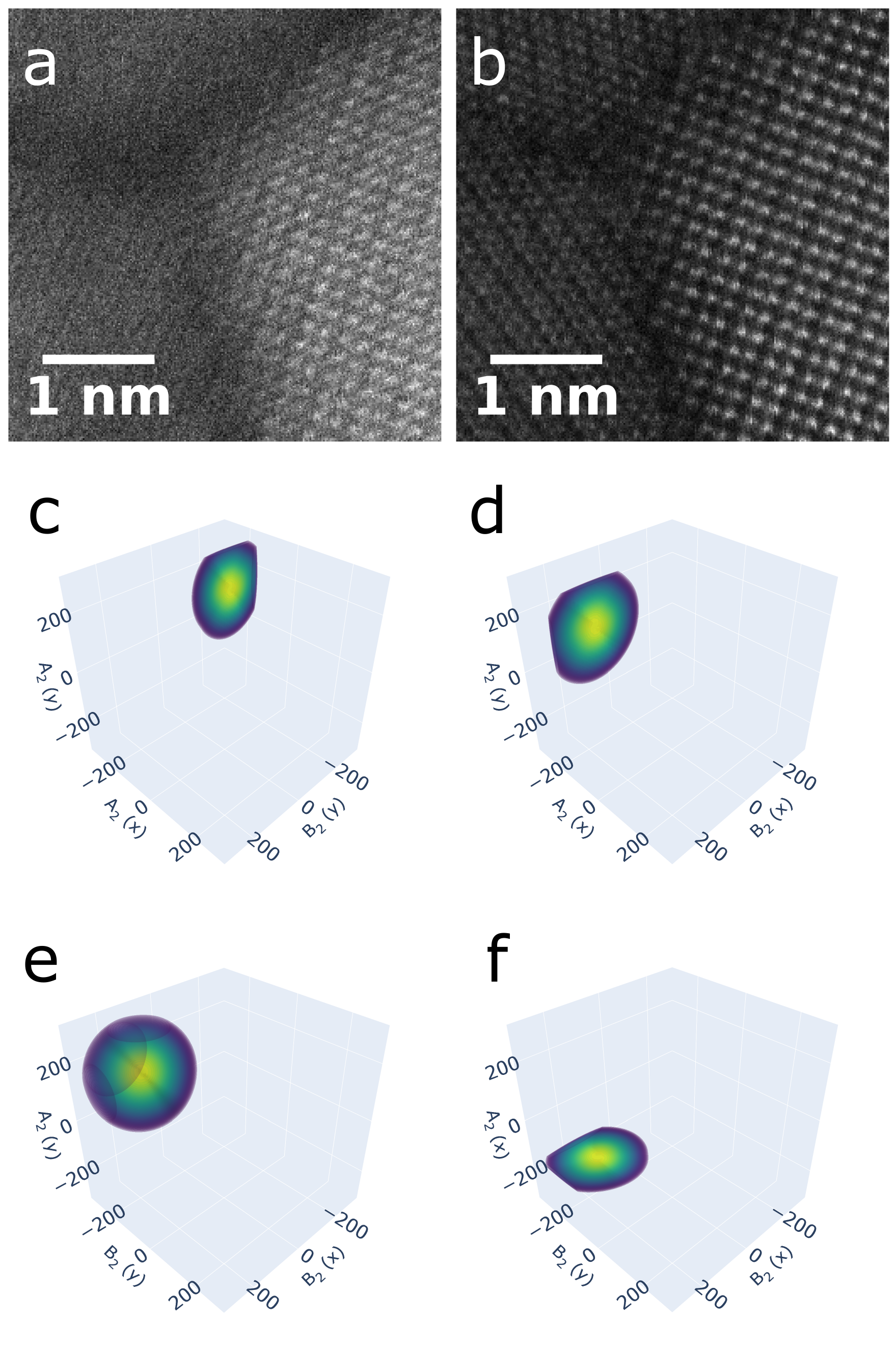}
\caption{BEACON correction of combined B\textsubscript{2} (x offset = 150 nm, y offset = -150 nm) and A\textsubscript{2} (x offset = -150 nm, y offset = 150 nm) aberrations on the HfO\textsubscript{2} thin film sample. Magnified portion of the full image a) before and b) after correction. c-f) Three-dimensional volumetric plots of the four-dimensional surrogate model. Each plot was generated using the optimum value for c) B\textsubscript{2,x}, d) B\textsubscript{2,y}, e) A\textsubscript{2,x}, f) A\textsubscript{2,y}.}
\label{fig:B2 A2 gpCAM}
\end{figure}

We grouped first- (C\textsubscript{1}, A\textsubscript{1}) and second-order (B\textsubscript{2}, A\textsubscript{2}) aberrations, because they affect image NV similarly simplifying the procedure.
Figure \ref{fig:C1 A1 gpCAM} shows the C\textsubscript{1}-A\textsubscript{1} optimization (three-dimensional) before (Figure \ref{fig:C1 A1 gpCAM}a) and after (Figure \ref{fig:C1 A1 gpCAM}c) 65 iterations using the HfO\textsubscript{2} thin film sample. 
Figure \ref{fig:C1 A1 gpCAM}b) shows the three-dimensional volumetric plot of the surrogate model of aberration space (estimated NV for each aberration set). 
A second optimization run for combined B\textsubscript{2}-A\textsubscript{2} with 85 iterations is shown in Figure \ref{fig:B2 A2 gpCAM}. 
The aberration space plots in Figure \ref{fig:B2 A2 gpCAM}c-f) are three-dimensional plots obtained by fixing one of the aberration coefficients to its optimum value, enabling the visualization of the four-dimensional model.
Equivalent low-magnification results using the Au NP sample are shown in Supplementary Figure \ref{fig:C1 A1 AuNP gpCAM} and Supplementary Figure \ref{fig:B2 A2 AuNP gpCAM}. 
The convergence results of multiple runs on both samples are presented in Table \ref{tab:combined statistics}.

\begin{table}
\centering
\begin{tabular}{|c|c|c|c|c|}
\hline
\begin{tabular}[c]{@{}c@{}}Corrected\\ aberration\end{tabular} & \begin{tabular}[c]{@{}c@{}}Precision of HfO$_2$\\ measurement (nm)\end{tabular} & \begin{tabular}[c]{@{}c@{}}Convergence \\ iteration\\ for HfO$_2$\end{tabular} & \begin{tabular}[c]{@{}c@{}}Precision of Au NP\\ measurement (nm)\end{tabular} & \begin{tabular}[c]{@{}c@{}}Convergence \\ iteration\\ for Au NP\end{tabular} \\ \hline
C\textsubscript{1}-A\textsubscript{1} & C\textsubscript{1}: 0.64, A\textsubscript{1}: 0.51 & C\textsubscript{1}: 57, A\textsubscript{1}: 57 & C\textsubscript{1}: 1.29, A\textsubscript{1}: 0.31 & C\textsubscript{1}: 20, A\textsubscript{1}: 20 \\ \hline
B\textsubscript{2}-A\textsubscript{2} & B\textsubscript{2}: 25.58, A\textsubscript{2}: 60.06 & B\textsubscript{2}: 69, A\textsubscript{2}: 49 & B\textsubscript{2}: 11.21, A\textsubscript{2}: 12.54 & B\textsubscript{2}: 36, A\textsubscript{2}: 71 \\ \hline
\end{tabular}
\caption{Precision and number of iterations required for convergence for combined first- and combined second-order aberration correction.}
\label{tab:combined statistics}
\end{table}

\begin{figure}
\centering
\includegraphics [width=\textwidth] {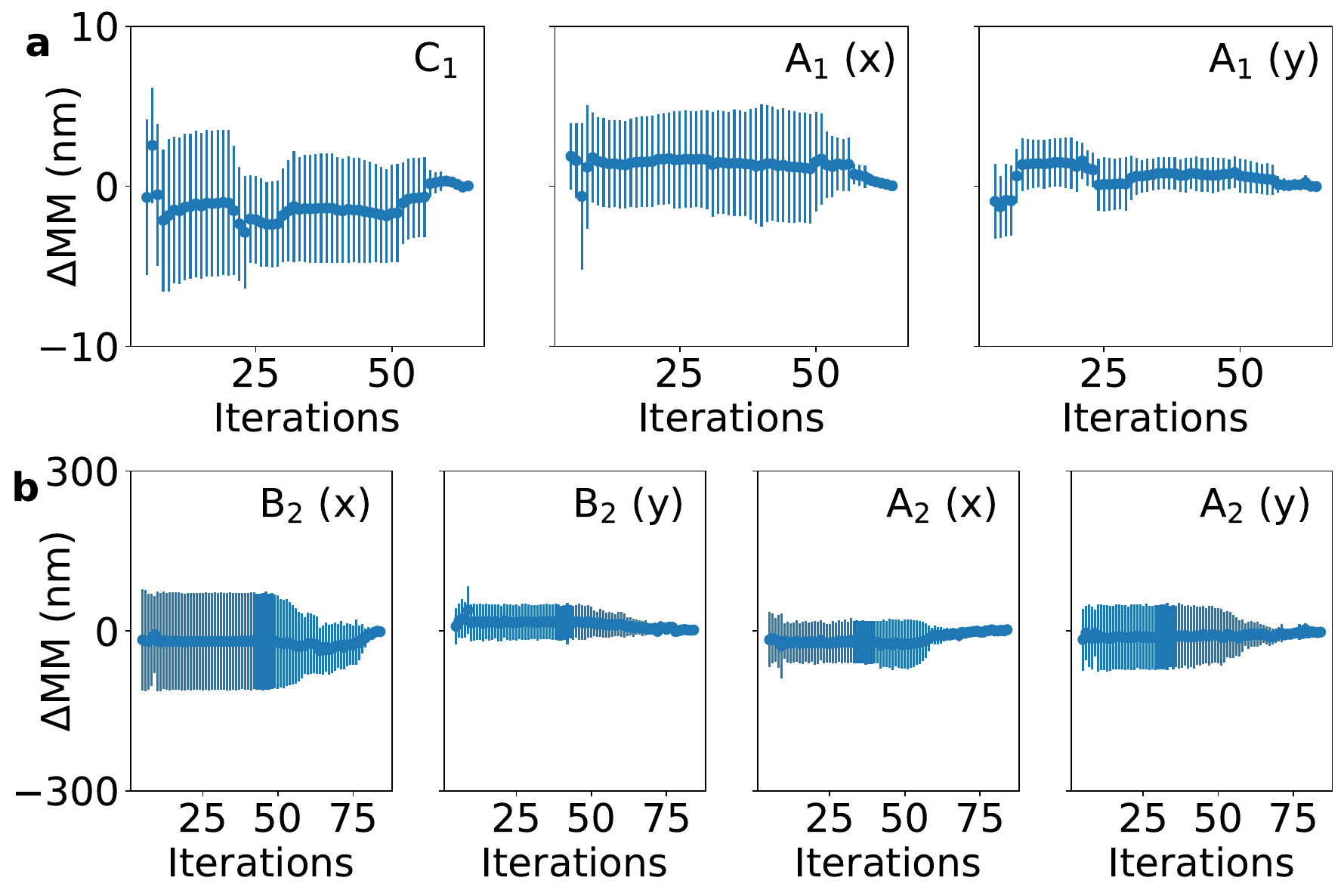}
\caption{Plots of the average difference between the model maximum and the final model maximum ($\Delta$MM) during optimization runs on HfO\textsubscript{2} thin film for a) C\textsubscript{1}-A\textsubscript{1} correction and b) B\textsubscript{2}-A\textsubscript{2} correction.}
\label{fig:combined convergences HfO2}
\end{figure}

\begin{figure}
\centering
\includegraphics [width=\textwidth] {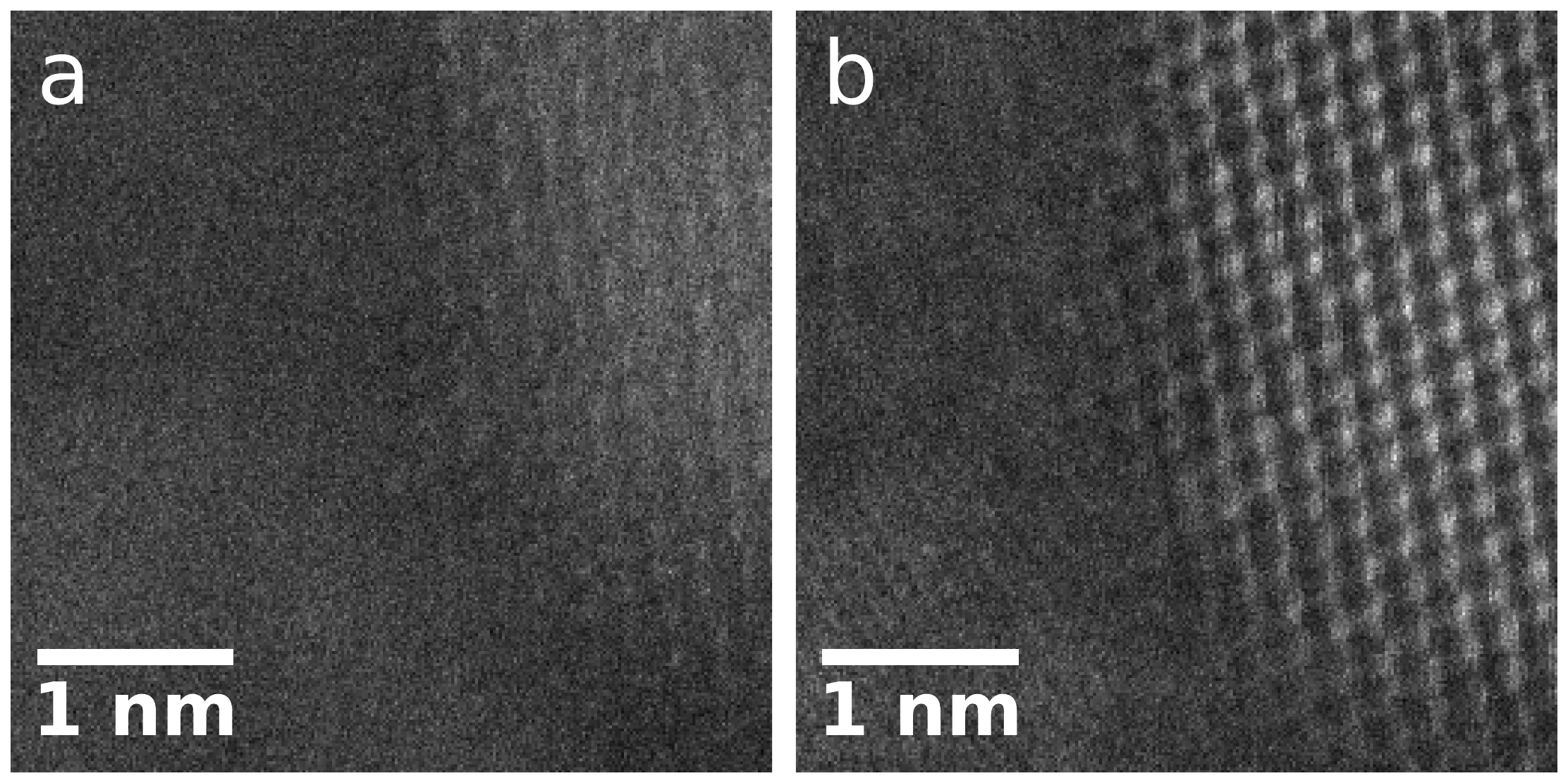}
\caption{BEACON correction of combined C\textsubscript{1} (offset = -5 nm), A\textsubscript{1} (x offset = -5 nm, y offset = 5 nm), B\textsubscript{2} (x offset = -150 nm, y offset = 150 nm) and A\textsubscript{2} (x offset = -150 nm, y offset = 150 nm) aberrations on HfO\textsubscript{2} thin film sample. Zoomed portion of the full image a) before and b) after correction.}
\label{fig:C1 A1 B2 A2 gpCAM}
\end{figure}

\begin{figure}
\centering
\includegraphics [width=\textwidth] {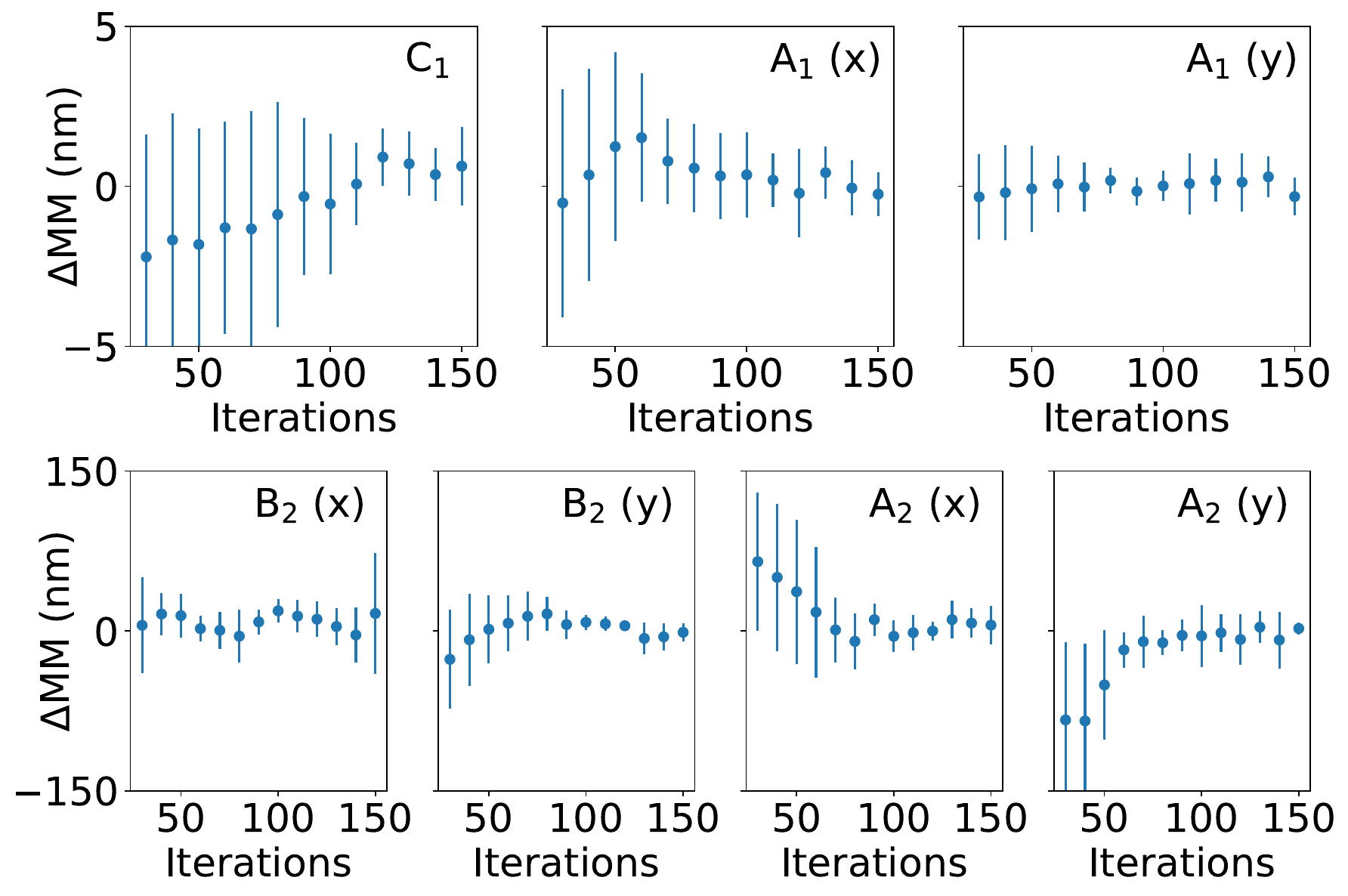}
\caption{Plots of average difference between model maximum and final model maximum ($\Delta$MM) during optimization runs on HfO\textsubscript{2} thin film for C\textsubscript{1}-A\textsubscript{1}-B\textsubscript{2}-A\textsubscript{2}.}
\label{fig:7D convergences HfO2}
\end{figure}

Finally, we optimized all first- and second-order aberrations simultaneously using 150 iterations. 
Figure \ref{fig:C1 A1 B2 A2 gpCAM} shows the results of these corrections on the HfO\textsubscript{2} thin film sample. 
Table \ref{tab:7D statistics} includes the precision of these corrections but does not list the number of iterations needed for convergence as none of the standard deviations of the differences from the final model maximum fell below our benchmarks. 
Figure \ref{fig:7D convergences HfO2} shows the differences between the model maxima and final model maxima. 
These were only saved every 10 iterations due to the large computation incurred to calculate the full 7-dimensional surrogate model.

\begin{table}
\centering
\begin{tabular}{|c|c|c|}
\hline
Corrected aberration & Precision for HfO\textsubscript{2} (nm) & Precision for Au NPs (nm) \\ \hline
C\textsubscript{1}-A\textsubscript{1}-B\textsubscript{2}-A\textsubscript{2} & \begin{tabular}[c]{@{}c@{}}C\textsubscript{1}: 4.44, A\textsubscript{1}: 1.17,\\ B\textsubscript{2}: 20.66, A\textsubscript{2}: 19.97\end{tabular} & \begin{tabular}[c]{@{}c@{}}C\textsubscript{1}: 5.10, A\textsubscript{1}: 9.36,\\ B\textsubscript{2}: 107.10, A\textsubscript{2}: 147.98\end{tabular} \\ \hline
\end{tabular}
\caption{Precision of C\textsubscript{1}-A\textsubscript{1}-B\textsubscript{2}-A\textsubscript{2} aberration correction.}
\label{tab:7D statistics}
\end{table}

\section{Discussion}\label{sec3}
\begin{figure}[H]
\centering
\includegraphics [width=\textwidth] {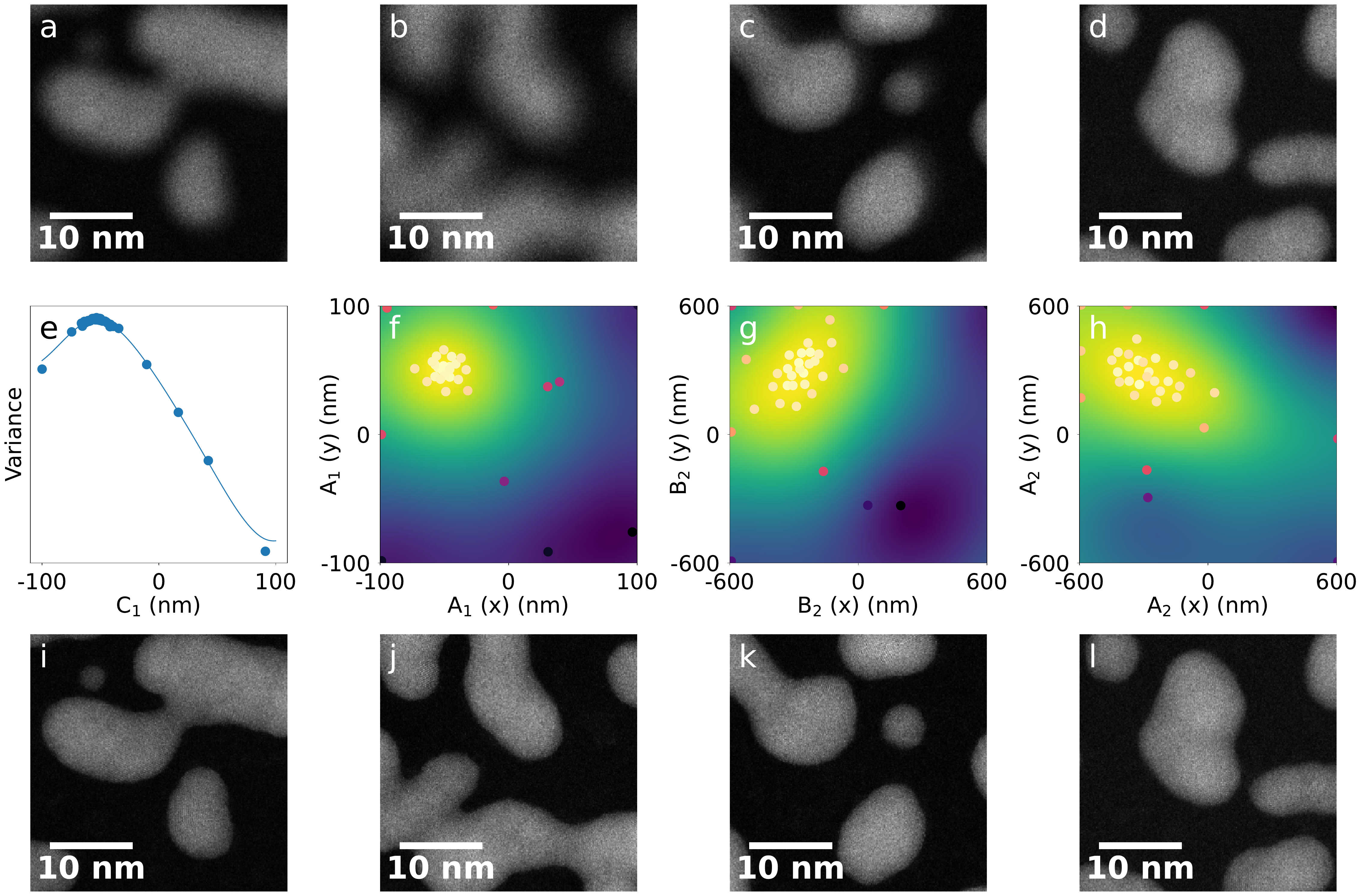}
\caption{Bayesian optimization of C\textsubscript{1} (offset = -50 nm) (a, e, i), A\textsubscript{1} (x offset = -75 nm, y offset = -75 nm) (b, f, j), B\textsubscript{2} (x offset = -300 nm, y offset = -300 nm) (c, g, k), and A\textsubscript{2} (x offset = -300 nm, y offset = -300 nm) (d, h, l) aberrations on the Au NP sample. a-d) Images before correction. e-h) Surrogate models with sampled points marked by dots and colored by NV. i-l) Images after correction. A zero aberration value was the initial well-corrected state as determined by the CEOS alignment software.}
\label{fig:AuNP gpCAM}
\end{figure}

\begin{figure}
\centering
\includegraphics [width=\textwidth] {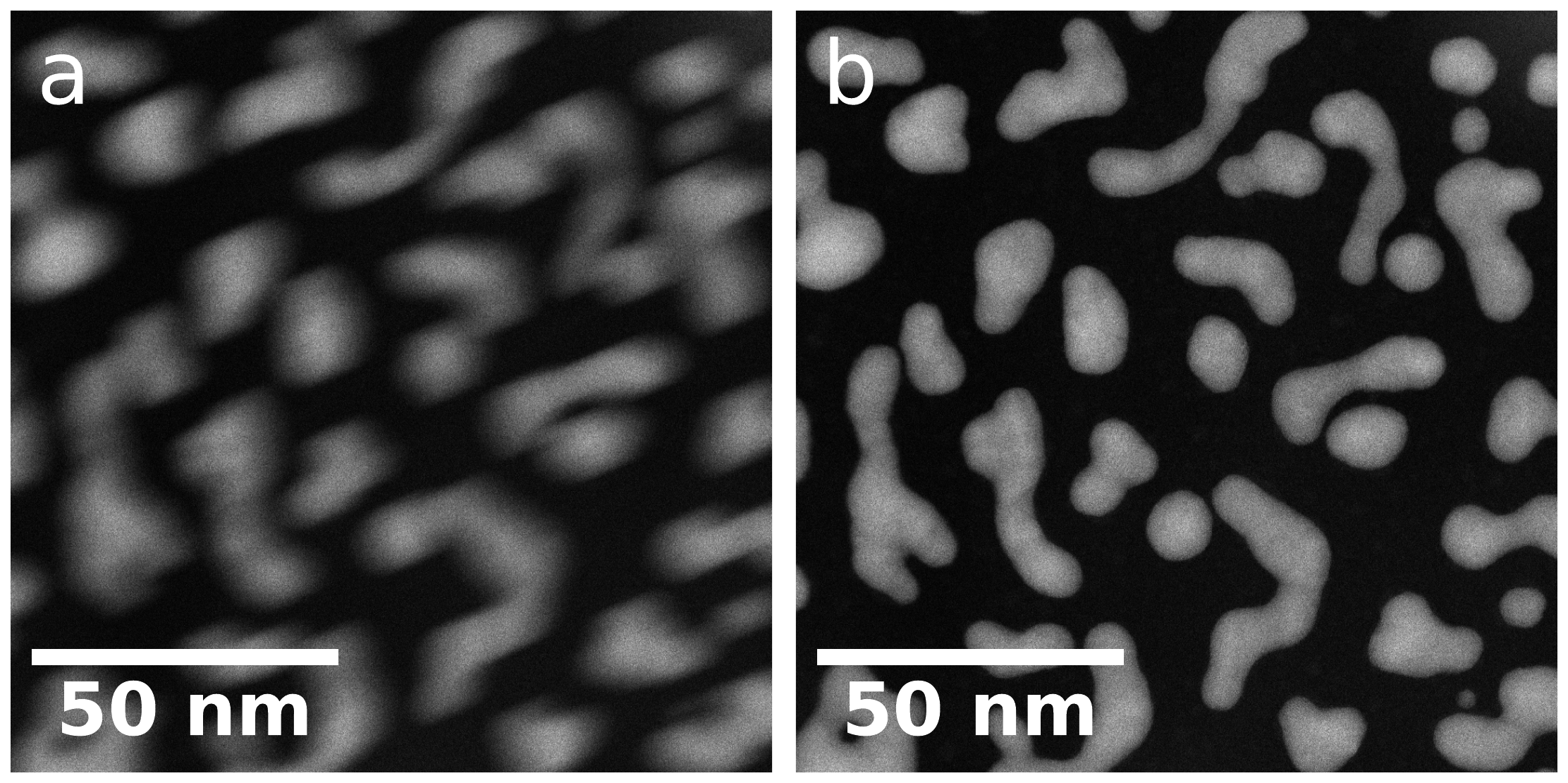}
\caption{Bayesian optimization of combined C\textsubscript{1} (offset = 50 nm) and A\textsubscript{1} (x offset = 50 nm, y offset = -50nm) aberrations on the Au NP sample. Image a) before and c) after correction.}
\label{fig:C1 A1 AuNP gpCAM}
\end{figure}

\begin{figure}
\centering
\includegraphics [width=\textwidth] {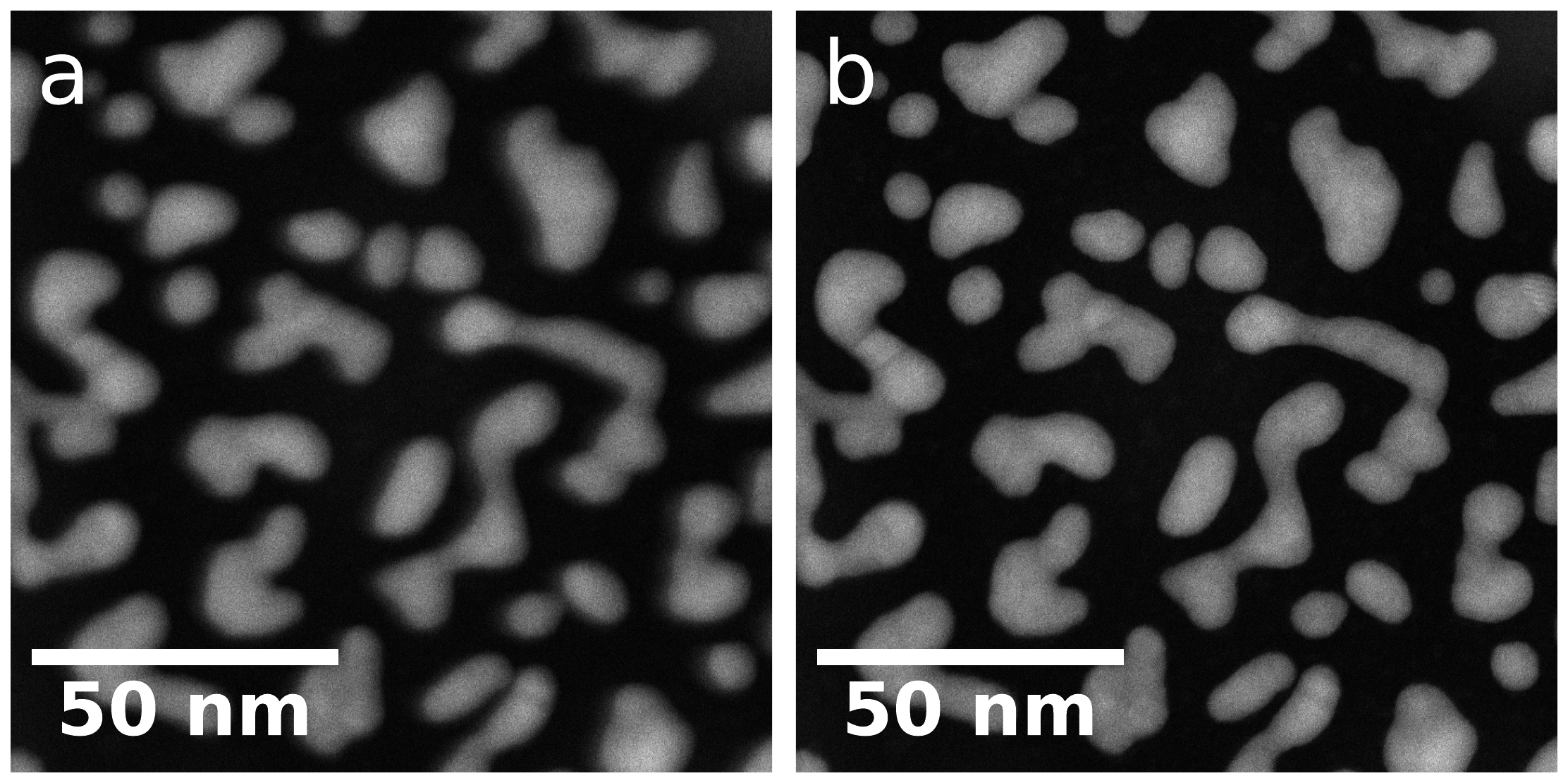}
\caption{Bayesian optimization of combined B\textsubscript{2} (x offset = 300 nm, y offset = -200 nm) and A\textsubscript{2} (x offset = -300 nm, y offset = 400 nm) aberrations on the Au NP sample. Image a) before and b) after correction.}
\label{fig:B2 A2 AuNP gpCAM}
\end{figure}

\begin{figure}
\centering
\includegraphics [width=\textwidth] {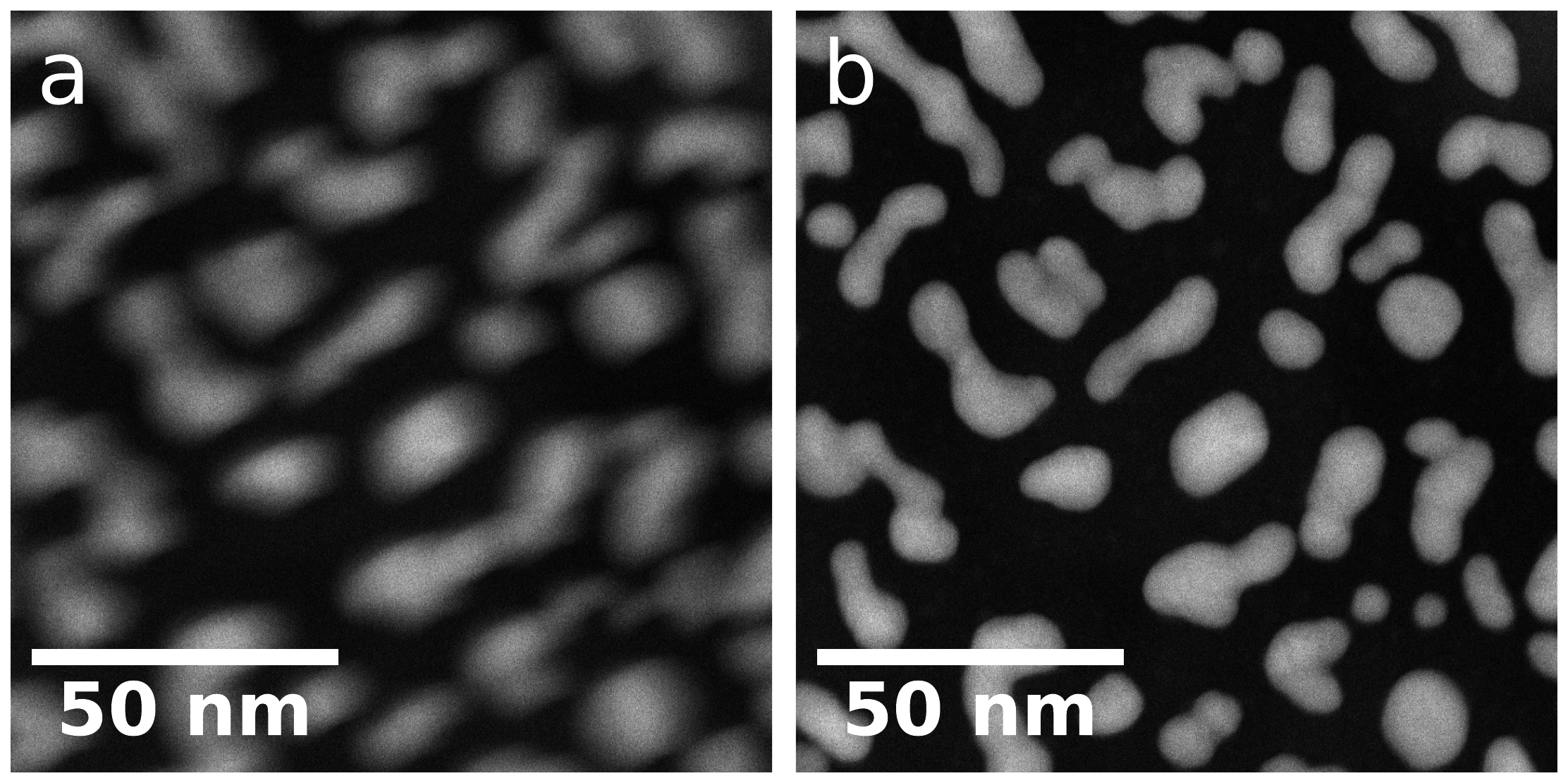}
\caption{Bayesian optimization of combined C\textsubscript{1} (offset = 0 nm), A\textsubscript{1} (x offset = -50 nm, y offset = 50 nm), B\textsubscript{2} (x offset = 300 nm, y offset = -200 nm) and A\textsubscript{2} (x offset = 300 nm, y offset = -400 nm) aberrations on Au NP sample. Image a) before and b) after correction.}
\label{fig:C1 A1 B2 A2 AuNP gpCAM}
\end{figure}

The before and after images in Figures \ref{fig:HfO2 gpCAM}, \ref{fig:C1 A1 gpCAM}, \ref{fig:B2 A2 gpCAM} and \ref{fig:C1 A1 B2 A2 gpCAM} show qualitative visual improvement in image quality for every optimization performed on the HfO\textsubscript{2} sample. 
Supplementary Figures \ref{fig:AuNP gpCAM}, \ref{fig:C1 A1 AuNP gpCAM}, \ref{fig:B2 A2 AuNP gpCAM} and \ref{fig:C1 A1 B2 A2 AuNP gpCAM} show similar improvements for the Au NP sample.

Table \ref{tab:individual statistics} indicates that optimizations of individual aberrations reached below the aforementioned benchmarks within a reasonable number of iterations. 
The one exception is that the precision of the A\textsubscript{2} correction on the Au NP sample was roughly double the benchmark. 
In our experience, this still constitutes an acceptable correction of A\textsubscript{2} which is typically corrected to $<$ 50 nm using the CEOS alignment software. 
With the exception of C\textsubscript{1}, convergence was slower for the HfO\textsubscript{2} as compared to the Au NP sample. 
We attribute this to the different noise levels, which can be seen by comparing the scatter of NV measurements for different C\textsubscript{1} values in Figure \ref{fig:HfO2 gpCAM} and Supplementary Figure \ref{fig:AuNP gpCAM}. 
We also note that the two samples required different noise functions (see Section \ref{sec:BEACON}). 
However, the number of iterations required to converge is relatively small for both samples, and BEACON achieves a corrected state with fewer images as compared to the traditional tableau method. 
A typical optimization run takes less than two minutes.
For comparison, acquisition of a single tableau using the CEOS alignment software used for correcting higher-order aberrations requires 44 images. Each tableau takes approximately one minute and multiple passes are often required to correct second-order aberrations to within our benchmarks.

Table \ref{tab:combined statistics} shows that the precision of the C\textsubscript{1}-A\textsubscript{1} corrections was similar to individual C\textsubscript{1} and A\textsubscript{1} corrections for both samples. 
The precision of B\textsubscript{2}-A\textsubscript{2} corrections is better than the individual B\textsubscript{2} and A\textsubscript{2} corrections for the Au NP sample but worse for the HfO\textsubscript{2}. 
In particular, the precision of the A\textsubscript{2} corrections is three times our benchmark. 
Additionally, the number of iterations required to converge is larger than the sum of the two individual aberrations in all cases except the C\textsubscript{1}-A\textsubscript{1} correction on the Au NPs. 
Here, the number of iterations required for C\textsubscript{1} and A\textsubscript{1} individually (21) is similar to the number required for the combination (20).

Table \ref{tab:7D statistics} shows that the precision of C\textsubscript{1}-A\textsubscript{1}-B\textsubscript{2}-A\textsubscript{2} corrections was comparable to individual optimizations for the HfO\textsubscript{2} thin film, with the exception of C\textsubscript{1}. 
However, the precision of corrections using this method on the Au NP sample is worse than corrections of each aberration individually or grouped by order. 
Furthermore, the standard deviation of the model maxima between different runs does not converge to below 1 nm for C\textsubscript{1} and A\textsubscript{1} or 20 nm for B\textsubscript{2} and A\textsubscript{2} for either sample. 
It is therefore more efficient to either correct the aberrations individually or grouped by order.

Overall, the results indicate that optimizing aberrations individually or grouped by order is more efficient and gives more precise corrections than optimizing all first- and second-order aberrations simultaneously. 
Still, correcting all first- and second-order aberrations collectively may be useful in certain circumstances. 
For example, as shown in Figure \ref{fig:7D convergences HfO2}, the combined C\textsubscript{1}-A\textsubscript{1}-B\textsubscript{2}-A\textsubscript{2} optimization successfully improves the alignment, but it fails to converge in a reasonable amount of time.
Such a coarse alignment could be followed by fine correction of each aberration individually. 
Additionally, further hyperparameter optimization may improve the performance of the combined correction method.

\begin{figure}
\centering
\includegraphics[width=\textwidth]{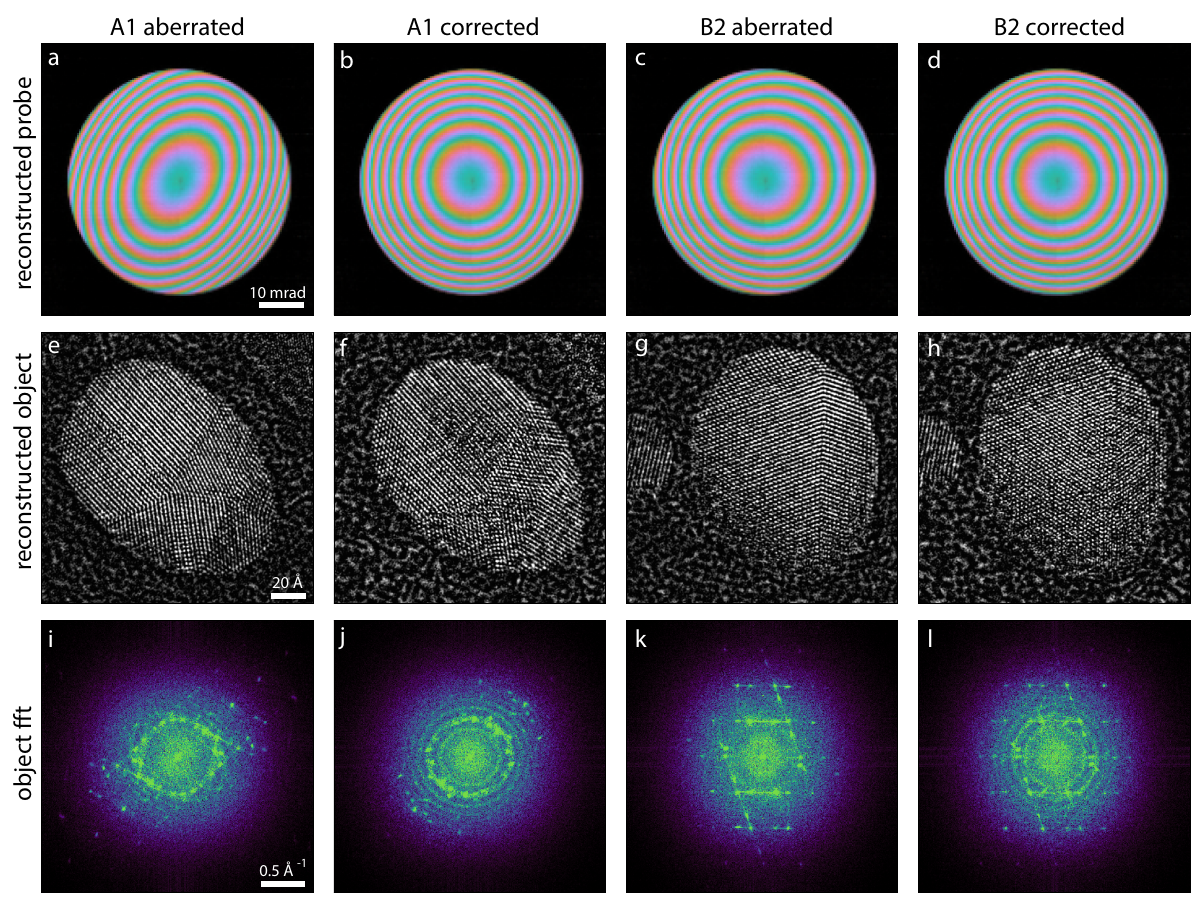}
\caption{Ptychographic reconstruction of two sets of particles: (a-d) probe in reciprocal space, (e-f) reconstructed object, and (g-h) object FFTs. Comparing a to b and c to d we observe reduction in aberrations after BEACON correction.}
\label{fig:ptycho}
\end{figure}

\subsection{Independent verification by electron ptychography}

As an independent verification of the BEACON method, we used ptychography to estimate aberrations from datasets with intentionally applied aberrations. 
Electron ptychography is a phase retrieval technique~\cite{Hoppe1969a,Hoppe1969b,Hoppe1969c,rodenburg1992,Nellist1995,chen2020mixed,varnavides2023iterative} that attempts to solve for both the object and the probe from a set of 2D diffraction patterns acquired at a set of probe positions, leading to the common term of 4D-STEM~\cite{Ophus2019-mx}.
Ptychographic reconstructions rely on careful control of the electron beam shape in order to acquire redundant information from the sample. 
Most often defocus is added to the incident probe to facilitate real space overlap between adjacent probe positions. 
It is also possible to correct for other aberrations based on the retrieved phase of the probe~\cite{nguyen2024achieving}. 

In this experiment, we added A\textsubscript{1} or B\textsubscript{2} to a well-corrected state before acquiring a 4D-STEM scan. 
We subsequently used BEACON to correct the aberrations and acquired a second dataset. We reconstructed both datasets using electron ptychography.
Although it is possible to solve for an unknown probe with iterative ptychography, reconstructions converge more quickly with a good initial guess, which includes aberrations. 
Using phase retrieval methods implemented in py4DSTEM, we typically estimate the probe aberrations with either a tilt-corrected BF-STEM (or parallax imaging) reconstruction~\cite{lupini2016,varnavides2023iterative,yu2024dose}, or using an independent Bayesian optimization algorithm~\cite{varnavides2023iterative}. 
The Bayesian optimization algorithm optimizes over self-consistency error of the ptychographic reconstruction, which in this case was used to estimate the initial aberrations in the probe. 
These fit aberration guesses provide an independent assessment of BEACON's ability to correct aberrations beyond image NV and visual image quality.

Figure \ref{fig:ptycho} shows the results of our reconstructions from ptychography experiments with Au NPs. 
Our initial probe for the first particle had large C\textsubscript{1} and A\textsubscript{1}, as illustrated in Figure \ref{fig:ptycho}a. 
The second column shows our second scan of the same particle after A\textsubscript{1} aberration correction by BEACON. 
Our estimate is that A\textsubscript{1} is reduced from 12 nm to 2 nm. 
The last two columns in Figure \ref{fig:ptycho} show a similar experiment with B\textsubscript{2} instead of A\textsubscript{1}. 
In this case, the probe begins with approximately 554 nm of B\textsubscript{2}, which is decreased to 224 nm. 
We note that aberration coefficients are defined differently by different groups and the CEOS coefficients used by BEACON are not expected to match the conventions used by py4DSTEM.
While large aberration values would interfere with conventional imaging, the ptychographic approach yields high-quality reconstructions with clearly resolved atomic features and an estimate of the probe shape for all scans.
Small differences between subsequent images of the sample particle can be due to beam damage and carbon contamination. 
Comparing the fast Fourier transforms (FFT) between these two scans (Figures \ref{fig:ptycho}i-j and \ref{fig:ptycho}k-l), we see similar spatial resolution but increased Thon ring contrast with the BEACON-corrected probes.

It has been suggested that increasing the phase diversity of the probe, through physical devices or aberration modulation, can improve information in ptychography reconstructions~\cite{yang2016enhanced, pelz2017low, ribet2023design, nguyen2024achieving}. 
However, it is often preferred to remove (especially unknown) higher-order aberrations from the incident beam. 
In this experiment, there was minimal impact on information transfer in the ptychographic reconstruction from the aberrated probe, but a clear reduction in the target aberration as evidenced by the final probe shape. 
Our ptychographic reconstructions highlight that BEACON lends itself especially well to the tuning of aberrations for many types of experiments, including conventional dark-field imaging and phase contrast experiments.

\subsection{BEACON performance considerations}\label{sec:performance}

Although this work demonstrates a reliable automated aberration correction method that works on different samples, several factors affect its behaviour and performance. 
Since BEACON relies solely on image NV, it is difficult to use in situations where image NV changes independent of aberration values. 
We found that imaging different sample regions (i.e. due to sample drift or beam shift) reduced the method's reliability. 
BEACON incorporates a method based on cross-correlation to ensure the same region is imaged in every iteration. 
Further, electron beam-induced carbon contamination also tends to reduce the sensitivity of the method. 
Reducing the applied dose by using a lower magnification or lower beam current can mitigate this, but a clean sample is always preferred.

Additionally, the sample geometry and specific region chosen can affect the behavior and performance. 
The Au NP sample has many sharp edges and relatively large image contrast. 
The image NV varies above the noise level over a broad range of aberration magnitudes at low magnifications. 
Other nanoparticle samples should behave similarly. 
At atomic resolution, small changes in the imaged region unexpectedly resulted in relatively large changes in NV. 
Normalization (see equation \ref{eq:NV}) is designed to compensate for this effect, but we found that using the same imaged region (identified by cross correlation) was an important feature to include in BEACON. 
Conversely, thin films such as the HfO$_2$ sample are less sensitive to this effect, because the intensity is relatively uniform across the entire sample. 
However, for this sample geometry the NV varies significantly only when atom columns or fringes are resolvable. 
Once atomic or lattice resolution is lost, NV is insensitive to aberration magnitude. 
Therefore, when the parameter bounds of the optimization are large and/or the optimization was performed far from the optimum condition, BEACON struggled to converge. 
In these cases, the NV peak is very narrow (approximating a Dirac delta function) compared to the parameter bounds. 
Tailoring the parameter bounds to approximately match the shape of the NV peak (which is sample-dependent) is essential to convergence.
Lastly, samples with low overall variance in intensity such as blank carbon films or amorphous materials probably cannot be used for this method.


Finally, the choice of noise estimation function was an important aspect of this work. 
Corrections performed on HfO\textsubscript{2} were relatively noisier when using the noise estimator designed for the Au NP sample. Thus, runs on the HfO$_2$ sample used equation \ref{eq:HfO2_noise} and runs on the Au NP sample used equation \ref{eq:AuNP_noise}. 
We are currently investigating the applicability of these noise functions to a range of different samples.

\subsection{Third-order aberrations}
Kirkland's simulations demonstrated that image NV can be used to optimize third-order aberrations: spherical aberration (C\textsubscript{3}), four-fold astigmatism (A\textsubscript{3}) and axial star (S\textsubscript{3})~\cite{kirkland2018fine}. 
Experimental verification proved difficult in our setup. 
Our software is unable to change C\textsubscript{3}, but we note that C\textsubscript{3} and A\textsubscript{1} are coupled. 
A combined C\textsubscript{3}-A\textsubscript{1} optimization could potentially correct both simultaneously essentially ``learning'' the relationship between the two aberrations. 
Also, changes to A\textsubscript{3} induced a large shift of the field-of-view which proved difficult to compensate for even with our cross-correlation method. 
The large image shifts necessary to recenter the field of view induce higher off-axial aberrations making the process self-defeating. 
The effect of S\textsubscript{3} on image NV is shown in Supplementary Figure \ref{fig:S3}, but the effect was so subtle that BEACON was less effective than the CEOS alignment software. 
Third-order aberrations tend to drift much less than first- and second-order aberrations during experiments, and we thus concentrated on the lower orders in this manuscript.

\section{Methods}\label{sec4}

\subsection{Normalized Variance (NV)}
Normalized variance (NV) is expressed as:

\begin{equation} \label{eq:NV}
    f(\boldsymbol{C_a}) = \frac{\sigma(\boldsymbol{C_a})^2}{\mu(\boldsymbol{C_a})^2} 
\end{equation}

\noindent
where $\boldsymbol{C_a}=[C_1,A_1,B_2,A_2,\cdots]$ is the vector or set of aberration coefficients, possibly scaled to the tolerance for each order. $\sigma$\textsuperscript{2} is the variance of the pixel intensities of an image, and $\mu$ is the mean pixel intensity, such that

\begin{equation}
    \mu = \frac{1}{N} \sum_{ij} z_{ij}(\boldsymbol{C_a})
\end{equation}

\begin{equation}
    \sigma^2 = \frac{1}{N} \sum_{ij} (z_{ij}(\boldsymbol{C_a})-\mu(\boldsymbol{C_a}))^2,
\end{equation}

\noindent
where \textit{z\textsubscript{ij}} is the pixel intensity at position \textit{i,j} and \textit{N} is the total number of pixels~\cite{kirkland2018fine}. Normalization is designed to reduce the sensitivity to specimen drift and changes in beam current. We note that the level of noise affects the image NV but does not vary with changes in aberrations. A greater concern is the build up of carbon during STEM scanning, which decreases the image NV. This can be mitigated by reducing dose by using shorter dwell times, fewer scan points, and minimizing the number of optimization iterations (i.e. the number of images acquired). SEM and HAADF-STEM are both essentially incoherent images and the NV behaves similarly. Traditional BF-CTEM is primarily coherent phase contrast, and NV has not been found to behave as described below. However, incoherent BF-CTEM most likely behaves similarly to HAADF-STEM.

\subsection{Bayesian Optimization}\label{BO}
Bayesian optimization is a quick and efficient method of estimating expensive-to-compute unknown functions by minimizing uncertainty~\cite{shahriari2015taking}. 
It generates a surrogate model as a proxy for the unknown function, usually using Gaussian processes (GP)~\cite{shahriari2015taking}. 
If we interpret data as a set of noisy function evaluations of a data-generating, inaccessible, ground-truth latent function \textit{f}($\boldsymbol{C_a}$), a GP assumes that a prior normal distribution can be placed over every finite subset of those function evaluations. 
This prior normal is fully defined by a prior mean and a covariance or kernel function~\cite{noack2022advanced}. 
This prior distribution can be optimized (trained) by maximizing the log marginal likelihood of the observed data. 
Once training is completed, conditioning on observations yields a posterior probability distribution for the latent function values across the domain. 
This posterior distribution provides a posterior mean (often considered the surrogate model) and posterior variances (often considered the uncertainties), both of which are used by the acquisition function such that maxima represent optimal new points for data collection. 
The choice of acquisition function defines the behaviour of the optimization routine~\cite{shahriari2015taking}. 
One commonly used acquisition function for locating the maximum of an unknown function is the upper confidence bound (UCB) method, where the next measurement point is chosen by the maximum of the UCB. 
This is mathematically represented by:

\begin{equation}
    \boldsymbol{C_a}^* = \argmax_{\boldsymbol{C_a}}{[m_{GP}(\boldsymbol{C_a})+c\sigma_{GP}(\boldsymbol{C_a})]},
\end{equation}

where $\boldsymbol{C_a}^*$ is the next sample point in aberration space, $m_{GP}(\boldsymbol{C_a})$ is the GP posterior mean, which is the predicted image quality (i.e. the image quality at aberration setting $\boldsymbol{C_a}$), $\sigma_{GP}(\boldsymbol{C_a})$ is the posterior uncertainty of the posterior image quality at aberration setting $\boldsymbol{C_a}$, and \textit{c} is a coefficient that determines the size of the confidence bound~\cite{shahriari2015taking}. 
Increasing \textit{c} favors exploration (searching new regions for a higher maximum) while decreasing \textit{c} favors exploitation (searching closer to the current maximum for a better estimate)~\cite{Berk2021}. 
Other acquisition functions for Bayesian optimization, such as predicted improvement and expected improvement~\cite{Gan2021}, can also be used, but we have found UCB to be most effective.
Here, we use Bayesian optimization to determine the optimum aberration corrector alignment. The latent function for the GP framework is the image NV, which we maximize to correct aberrations.

\subsection{BEACON operation}\label{sec:BEACON}
\begin{figure}
\centering
\includegraphics [width=\textwidth] {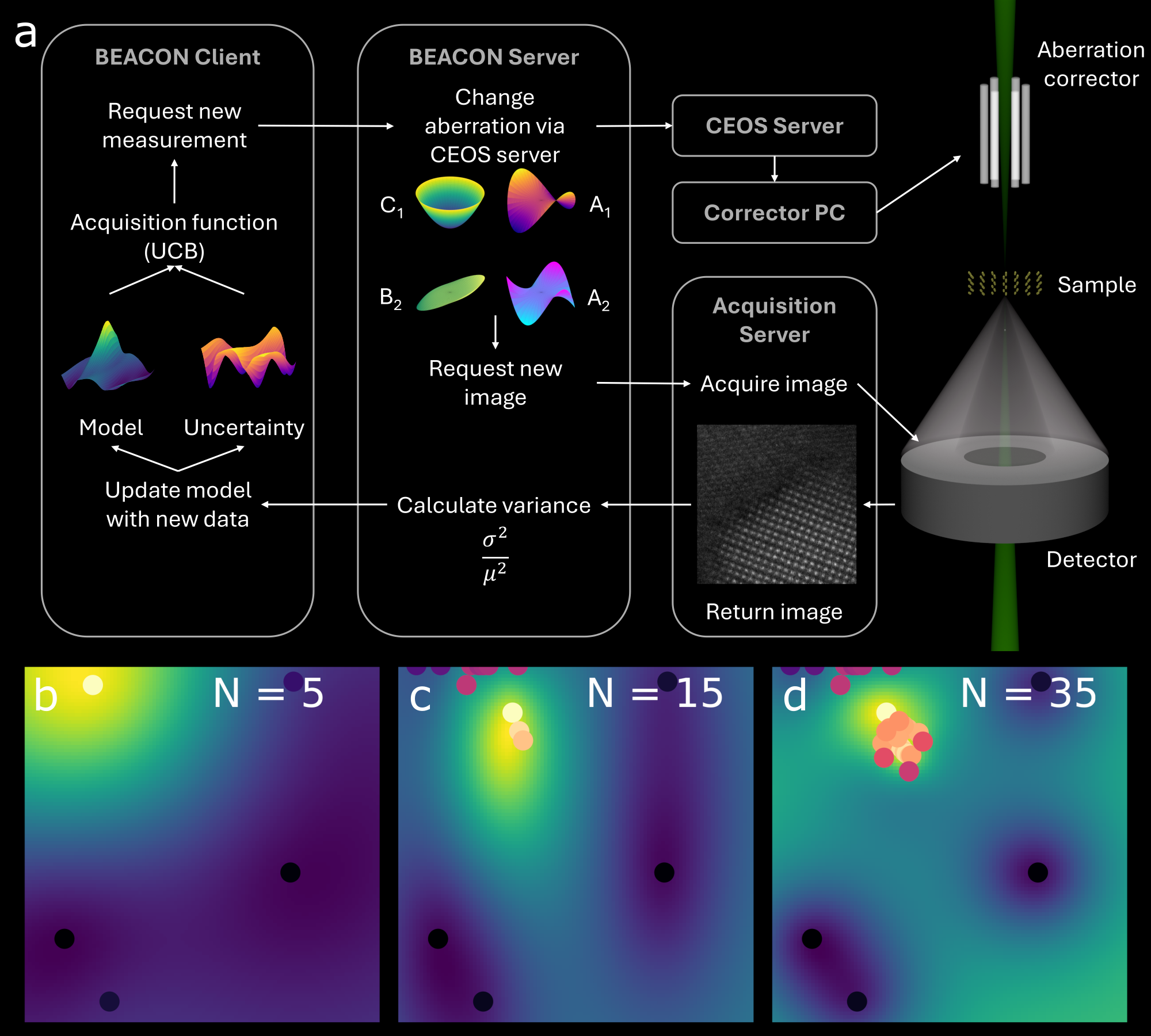}
\caption{a) Software schematic of aberration correction workflow. b-d) Evolution of an A\textsubscript{1} surrogate model after b) 5 iterations, c) 15 iterations and d) 35 iterations.}
\label{fig:methods figure}
\end{figure}

As shown in Figure \ref{fig:methods figure}a, BEACON uses a client-server model to separate the Bayesian optimization process from the microscope control computer.
First, the client requests a new measurement with a specific set of aberration coefficients. 
The server changes the state of the aberration corrector and acquires a new HAADF-STEM image. 
The server then calculates the image NV and returns this value to the client. The acquired image can also be returned to the client.
The client updates the surrogate model of image quality for all sets of aberration coefficients as well as the model uncertainty. 
The UCB acquisition function then uses the surrogate model and the model uncertainty to choose the next set of aberration coefficients to measure.
Figure \ref{fig:methods figure}b-d show how the surrogate model evolves as more data points are added. 
The region of parameter space in which the peak resides is located with just 5 data points, as shown in Figure  \ref{fig:methods figure}b. 
Subsequent sampling efforts are focused in this region, and the approximate location of the peak is located with 15 data points, as shown in Figure \ref{fig:methods figure}c. 
BEACON continues to search around this peak to accurately identify the peak, as shown in Figure \ref{fig:methods figure}d.
This shows how BEACON can efficiently optimize the aberrations without needing to sample large, extraneous portions of parameter space.

We acquired 512 by 512-pixel images with a dwell time of 3 $\mu$s and a current of 30 pA. 
Large adjustments to aberrations, particularly B\textsubscript{2} and A\textsubscript{2}, result in beam shifts that change the field of view. 
We use cross-correlation to ensure the same 256 by 256 pixel image region is used throughout an optimization run which also compensates for sample drift. 
The field of view was 7.8 nm for the HfO\textsubscript{2} thin film and 62.4 nm for the Au NP sample. 
BEACON incorporates a graphical user interface (GUI) as shown in Supplementary Figure \ref{fig:GUI} that enables users to quickly tune the optimization parameters for their sample. This GUI is included in the BEACON software, which is generally available (see Code Availability for details).

Bayesian optimization can incorporate noise variances into the calculation of surrogate models.
Initial testing showed that proper noise estimation was vital to both the speed and precision of optimization.
Measurements from the HfO\textsubscript{2} sample were noisier than those from the Au NP sample, requiring the use of different noise estimation functions. 
For the HfO\textsubscript{2} sample, the noise function \textit{S} was:

\begin{equation} \label{eq:HfO2_noise}
    S = 0.0001 \times (y_{max}-y_{min}).
\end{equation}

For the Au NP sample, we found that the default noise estimation function used by gpCAM,

\begin{equation} \label{eq:AuNP_noise}
    S = 0.01 \times \lvert\Bar{y}\rvert,
\end{equation}

was effective.

\subsection{Ptychographic reconstructions}
For the ptychography experiments, scans were collected at 300 kV using a probe defined with a 25 mrad convergence angle. Approximately 400 \AA~of defocus was added, and the step size of the acquisition was 1.4 \AA. 
A dose of approximately $2\times10^4$e$^-$/\AA $^2$ was used. 
We used the 4D Camera to acquire scans~\cite{Ercius2024-ed}. 
We reconstructed both datasets using a gradient descent algorithm with batched updates implemented in the open-source toolkit \texttt{py4DSTEM v 0.14.8} ~\cite{savitzky2021py4dstem, varnavides2023iterative}. 
A batch size of 1024 measurements was used (1/16 the total number of probe positions), and 100 iterations were run to converge the mean squared error. 
The probe amplitude was fixed using a vacuum probe measurement taken during the same session. 
Position refinement and slight low pass filtering was applied during each update. 
The initial probe aberrations were estimated from parallax reconstructions and Bayesian optimization algorithms. 
The defocus was estimated from a parallax reconstruction. 
The A\textsubscript{1} or B\textsubscript{2} coefficients were estimated from Bayesian optimization minimizing the mean squared error with 25 initial, uniformly spaced points followed by 25 more optimized calls.  
More details about the reconstruction formalism can be found elsewhere~\cite{varnavides2023iterative}.

\section*{Data Availability}
The data used to generate figures \ref{fig:HfO2 Grid Search} and \ref{fig:AuNP Grid Search} is available from the Zenodo repository (DOI: 10.5281/zenodo.13891904). Other data is available upon request.

\section*{Code Availability}
The BEACON code is available from our Github repository (\url{https://github.com/AlexanderJPattison/BEACON}).

\bibliography{BEACON-References}

\section*{Acknowledgements}
This work was primarily funded by the US Department of Energy in the program ``Electron Distillery 2.0: Massive Electron Microscopy Data to Useful Information with AI/ML." 
Work at the Molecular Foundry was supported by the Office of Science, Office of Basic Energy Sciences, of the U.S. Department of Energy under Contract No. DE-AC02-05CH11231. 
This research used resources of the National Energy Research Scientific Computing Center (NERSC), a Department of Energy Office of Science User Facility using NERSC award BES-ERCAP0028035. 
JP acknowledges financial support from the National Research Foundation of Korea (NRF) grant, funded by the Korea government (MSIT) (Grant No. RS-2023-00283902, and RS-202400408823).
We gratefully acknowledge CEOS, GmbH for providing the server enabling communication with the corrector and I. Massmann for technical assistance.

\section*{Author Information}
\subsection*{Author Contributions}
A.J.P. developed the BEACON software and assessed its performance. S.M.R. verified BEACON performance using ptychographic reconstructions. M.M.N. is the developer of gpCAM and assisted with the development of BEACON. A.J.P., S.M.R., M.M.N., and P.E. wrote the draft manuscript. S.M.R., G.V. and C.O. are developers of py4DSTEM. K.P. and J.P. created the HfO\textsubscript{2} thin film sample. E.J.K. is the developer of the image NV theory and provided helpful discussions. P.E. supervised the project and assisted with conceptualization. All authors have read and approved the manuscript. 

\section*{Ethics Declarations}
\subsection*{Competing Interests}
All authors declare no financial or non-financial competing interests. 

\newpage

\section*{Supplementary information}

\begin{figure}
\centering
\includegraphics [width=\textwidth] {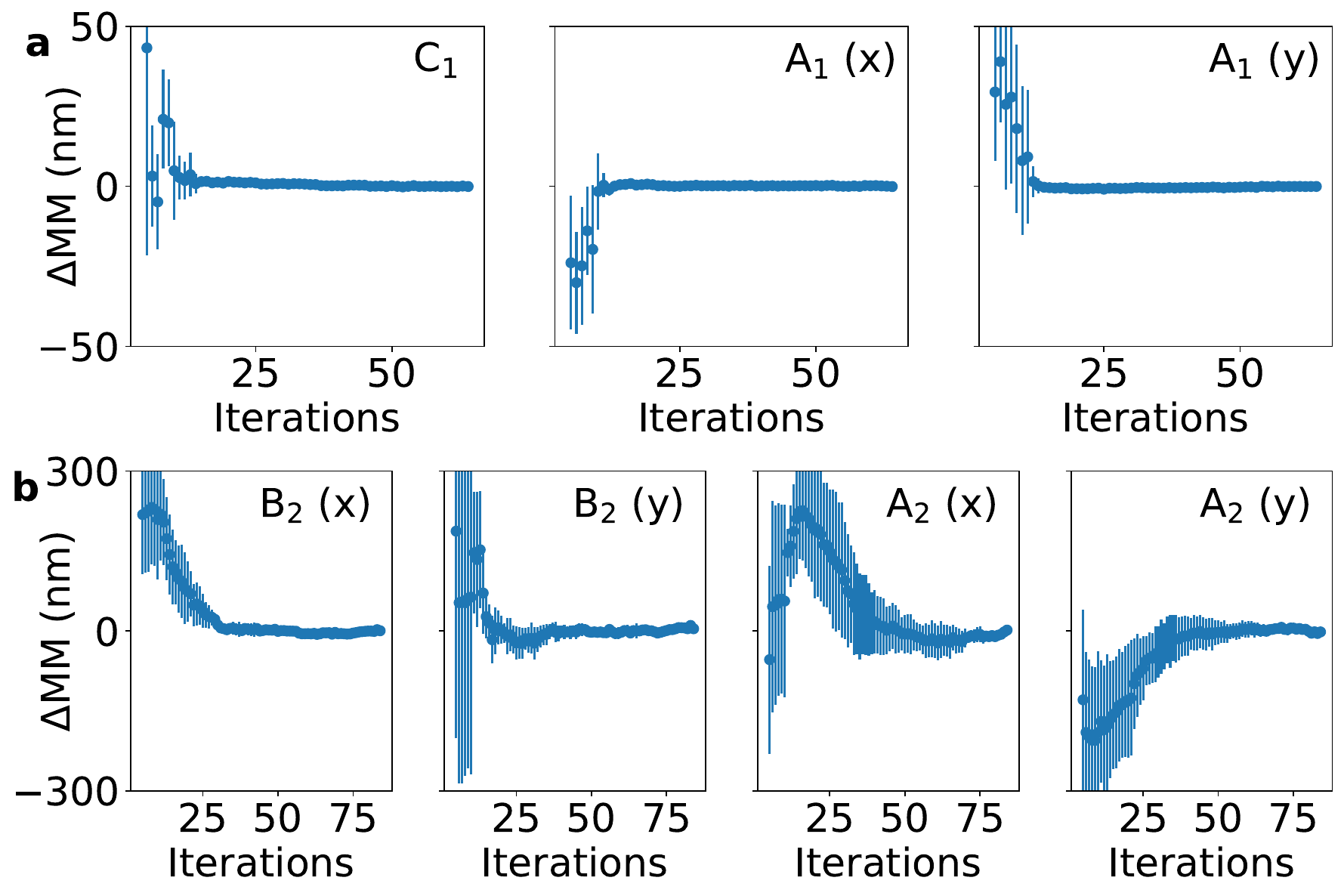}
\caption{Plots of average difference between model maximum and final model maximum ($\Delta$MM) during BEACON runs on the Au NP sample for a) C\textsubscript{1}-A\textsubscript{1} correction and b) B\textsubscript{2}-A\textsubscript{2} correction.}
\label{fig:combined convergences AuNP}
\end{figure}

\begin{figure}
\centering
\includegraphics [width=\textwidth] {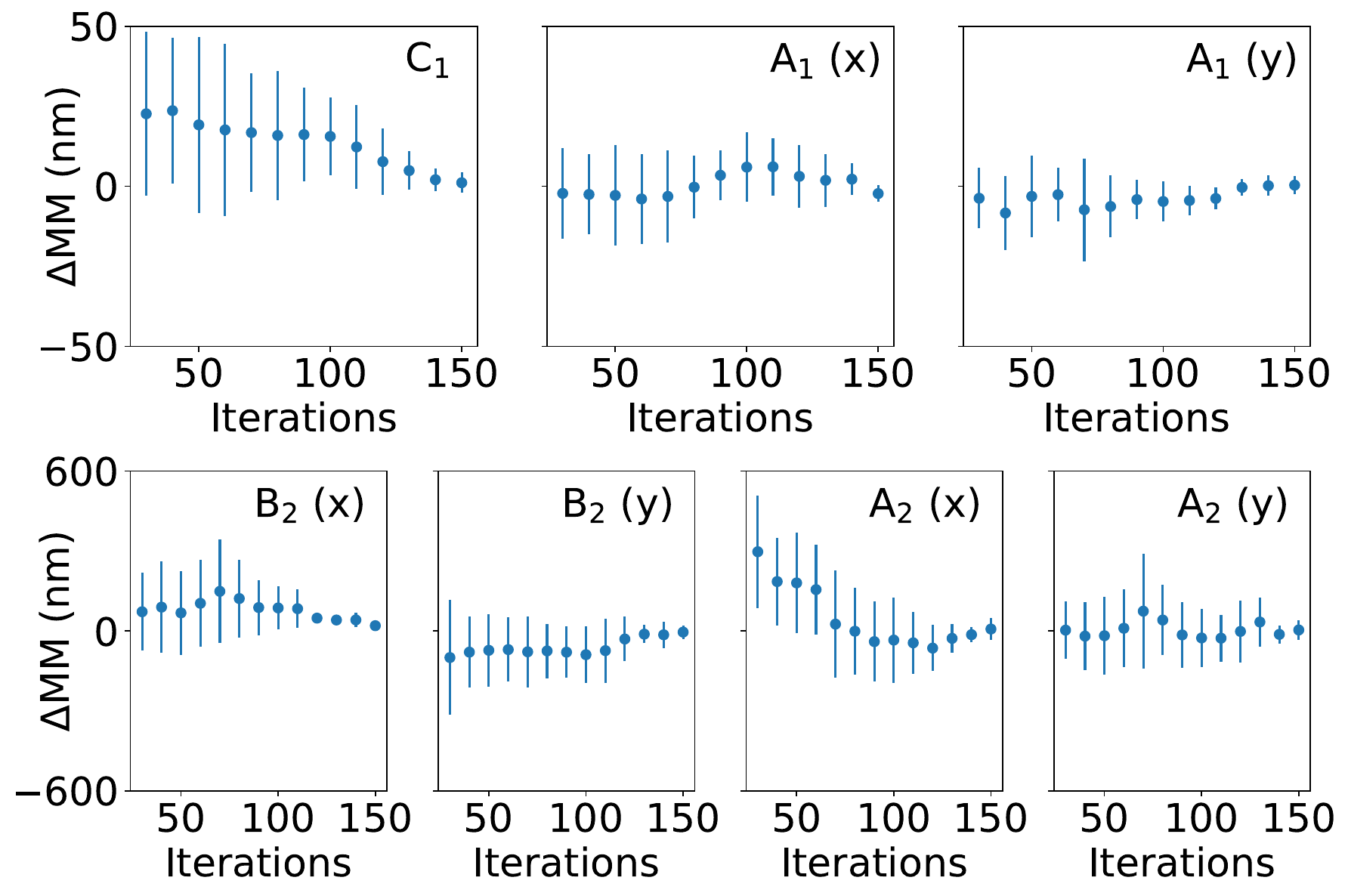}
\caption{Plots of average difference between model maximum and final model maximum ($\Delta$MM) during BEACON runs on the Au NP sample for C\textsubscript{1}-A\textsubscript{1}-B\textsubscript{2}-A\textsubscript{2}.}
\label{fig:7D convergences AuNP}
\end{figure}

\begin{figure}
\centering
\includegraphics [width=\textwidth] {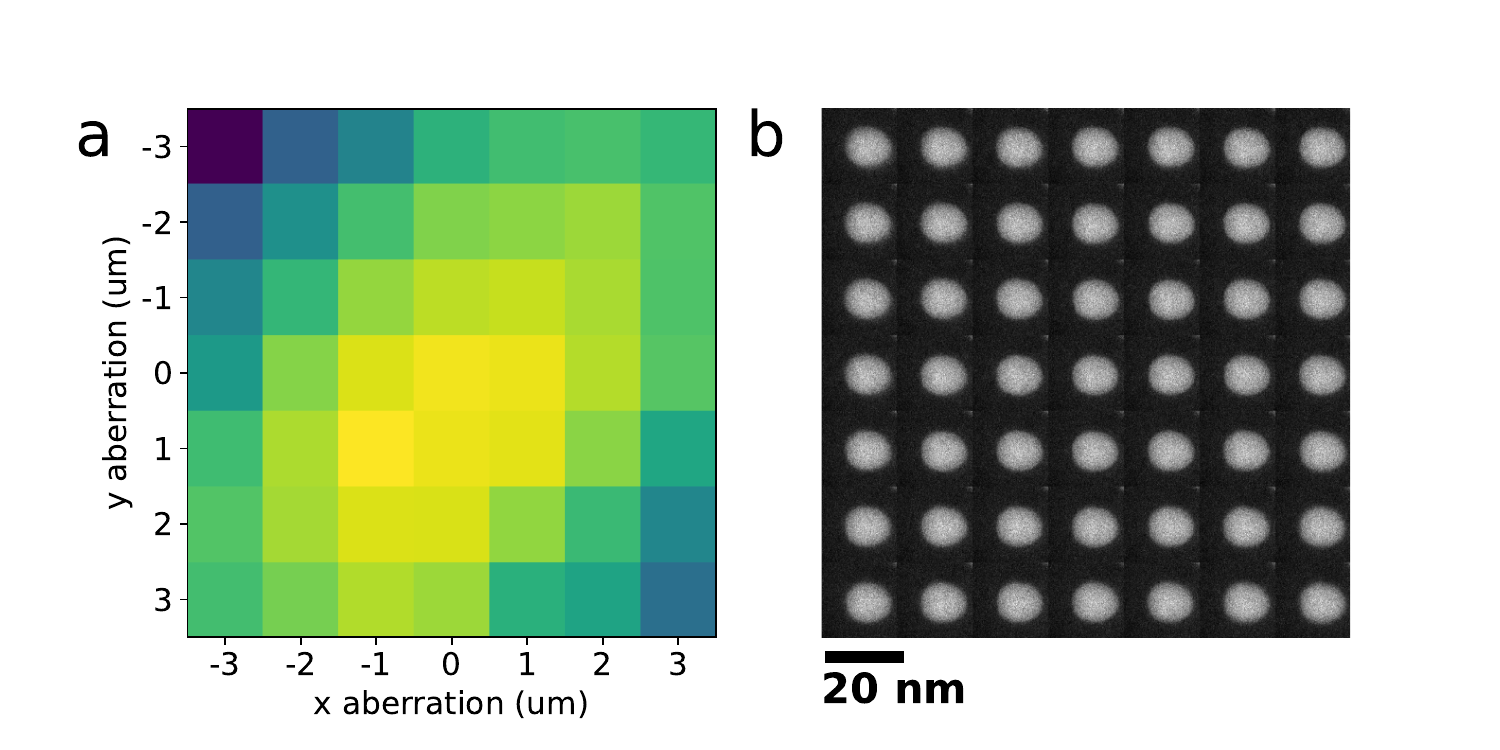}
\caption{a) Image NV for a grid search using the Au NP sample and b) tiled cutouts of the real space images (field of view = 24.4 nm) for the corresponding aberration states for S\textsubscript{3}.}
\label{fig:S3}
\end{figure}

\begin{figure}
\centering
\includegraphics [width=\textwidth] {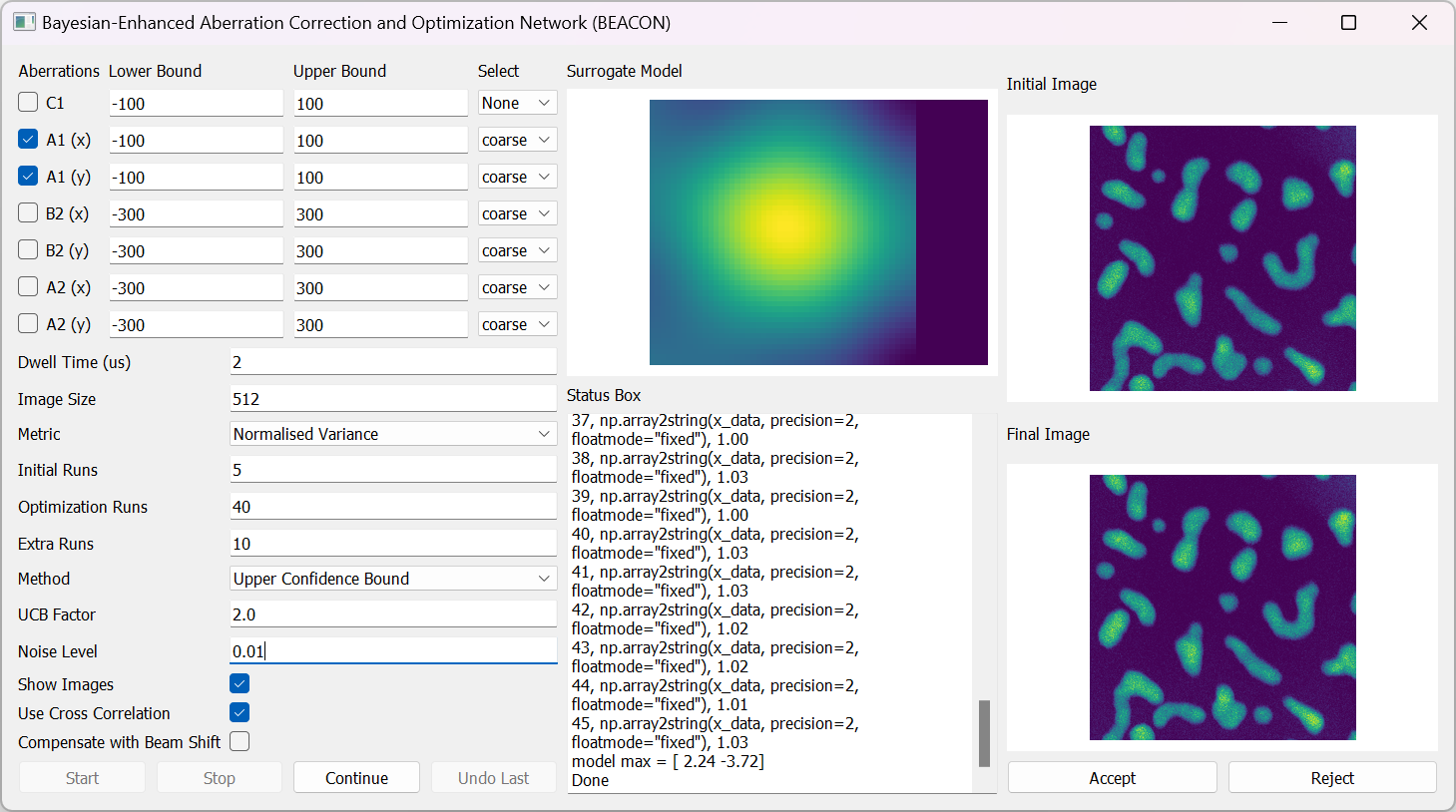}
\caption{BEACON graphical user interface. }
\label{fig:GUI}
\end{figure}

\end{document}